\documentclass[12pt]{article}

\usepackage[letterpaper, margin=1in]{geometry}
\usepackage[T1]{fontenc}
\usepackage{graphicx}
\usepackage{amsmath,amssymb}
\usepackage[table]{xcolor}
\usepackage{tabularx,multirow,  array}
\usepackage{hyperref}
\usepackage{enumitem}
\usepackage{subcaption}
\usepackage[normalem]{ulem}
\usepackage{cancel}
\usepackage{cite}
\usepackage[affil-it]{authblk}

\begin{document}

\title{Coupling Spectrum Estimation and Single Energy Material Decomposition via X-ray Grating Interferometry}

\author[1,2]{Longchao Men}
\author[1,2]{Peiyuan Guo}
\author[1,2]{Jincheng Lu}
\author[3]{Yan Xu}
\author[3]{Hongxia Yin}
\author[3]{Zhenchang Wang}
\author[1,2]{Li Zhang}
\author[1,2,*]{Zhentian Wang}

\affil[1]{Department of Engineering Physics, Tsinghua University, Beijing 100084, China}
\affil[2]{Key Laboratory of Particle \& Radiation Imaging (Tsinghua University) of Ministry of Education, Beijing 100084, China}
\affil[3]{Department of Radiology, Beijing Friendship Hospital, Capital Medical University, Beijing 100069, China}
\affil[*]{Corresponding author: wangzhentian@tsinghua.edu.cn}

\date{}

\maketitle

\begin{abstract}
Dark-field imaging based on grating interferometry is an emerging X-ray modality in medical imaging, which is particularly capable of providing complementary diagnostic information by visualizing the microstructural properties of lung tissue. However, quantitative dark-field imaging remains fundamentally challenged by beam hardening, which arises from the energy-dependent fringe visibility under polychromatic illumination. The resulting artifacts substantially degrade the quantitative accuracy of dark-field images.
In this work, motivated by our key observation of an intrinsic similarity between the X-ray energy spectrum and the system-related coupling spectrum, we propose a unified framework to simultaneously and independently estimate both spectra. By measuring the transmission associated with the zeroth- and first-order components of the phase-stepping curve using solid step-wedge phantoms, the two spectra are robustly estimated via an expectation-maximization algorithm. The recovered spectra are subsequently incorporated into a physics-based correction model to mitigate beam-hardening-induced artifacts in dark-field imaging effectively.
Furthermore, leveraging the inherent availability of two independent spectra within X-ray grating interferometry, we introduce a single-energy material decomposition method that achieves basis material imaging without dual-energy scans. Wave-optical simulations and experiments demonstrate accurate spectrum estimation, effective dark-field signal correction, and reliable material decomposition.
Consequently, the proposed framework extends the diagnostic potential of X-ray grating interferometry beyond pulmonary imaging, facilitating broader applications in medical imaging.
\end{abstract}


\section{Introduction}
\label{sec:introduction}
X-ray dark-field imaging (DFI) is a highly promising medical imaging modality, which originates from X-ray small-angle scattering generated by interactions with sample microstructures\cite{stroblGeneralSolutionQuantitative2014a, lynchInterpretationDarkfieldContrast2011, yashiroOriginVisibilityContrast2010}. It enables the characterization of microstructural information at scales much smaller than the spatial resolution of imaging system. Compared to conventional absorption imaging, dark-field imaging exhibits higher sensitivity to weakly absorbing objects with microstructures, superior image contrast.
Building upon this distinctive property, DFI has found applications in multiple medical fields. These include investigating microstructural changes in bone due to osteoporosis\cite{rischewskiDarkfieldRadiographyDetection2024}, classifying types of kidney stones\cite{schererNoninvasiveDifferentiationKidney2015} and breast calcifications\cite{wangNoninvasiveClassificationMicrocalcifications2014}, and using microbubbles as contrast agents for image enhancement\cite{langMicrobubblesContrastAgent2019}.

A primary application of DFI is the detection of early-stage pulmonary diseases. This potential originates from the air-filled alveolar microstructure of the lungs, in which abundant air-tissue interfaces produce strong small-angle scattering, thereby generating a pronounced dark-field signal in healthy pulmonary tissue\cite{gassertXrayDarkFieldChest2021}. When disease disrupts this microstructure, the dark-field signal is correspondingly reduced.
Clinical studies have continued to highlight the diagnostic value of dark-field imaging across a range of pulmonary conditions, including emphysema resulting from alveolar rupture\cite{linDetectionEarlyPulmonary2024, urbanDarkFieldChestRadiography2023}, fibrosis\cite{hellbachXrayDarkfieldRadiography2017}, lung cancer\cite{guoGratingbasedXrayDarkfield2024} and pneumothorax\cite{hellbachDepictionPneumothoracesLarge2018, gassertDarkfieldChestRadiography2025}.
The transition from laboratory to clinical settings has been marked by the TUM group's establishment of human-scale dark-field chest radiography\cite{willerXrayDarkfieldChest2021} and computed tomography (CT) systems\cite{viermetzDarkfieldComputedTomography2022}. Meanwhile, dedicated optimization strategies for pulmonary disease imaging in dark-field CT systems have been developed to further advance this technology toward clinical translation\cite{guoOptimizationXrayDarkfield2025, spindlerSimulationStudyXray2025}.

X-ray grating interferometry (XGI) stands as one of the most widely employed techniques for realizing DFI under laboratory conditions\cite{pfeifferHardXrayDarkfieldImaging2008}. The dark-field signal is quantitatively defined as the reduction in visibility of the periodic interference fringes generated via grating modulation. However, under polychromatic X-ray illumination, the fringes' visibility is influenced not only by small-angle scattering but also by the energy spectra. Although solid objects lacking microstructures do not generate small-angle scattering, the effect of beam hardening may decrease fringe visibility, consequently resulting in a dark-field signal that is purely induced by the hardening of the spectral\cite{yashiroEffectBeamHardening2015}.
This effect can significantly compromise the accuracy of dark-field imaging for representing microstructural information, particularly in clinical applications such as pulmonary imaging where the thoracic cavity contains various dense structures (e.g., bones and blood vessels) that can induce substantial beam hardening effects.
To address this challenge, various correction methods have been proposed. These include simulation-based approaches that incorporate energy-dependent attenuation into the dark-field signal model\cite{pelzerBeamHardeningDispersion2016, tangPixelwiseBeamhardeningCorrection2023, viermetzInitialCharacterizationDarkField2023} and empirical calibration techniques based on step-wedge phantoms\cite{lochschmidtLookupTablecorrectionBeam2025, kasterBeamHardeningCorrection2025}.

In this work, we first report a key observation that the visibility spectrum, which characterizes the energy dependence of fringe visibility, is intrinsically coupled with the X-ray energy spectrum, forming a unified coupling spectrum with strong structural similarity to the energy spectrum. Based on this finding, we develop a framework that enables the simultaneous yet independent estimation of the energy spectrum and the coupling spectrum using measurements acquired with solid step-wedge phantoms.
The retrieved spectra can be used to effectively mitigate beam-hardening-induced artifacts in dark-field imaging and potentially facilitate the correction of visibility spectrum hardening\cite{demarcoXRayDarkFieldSignal2024}.
Furthermore, leveraging the inherent availability of two independent spectra within a single interferometric acquisition, we introduce a method termed single-energy material decomposition (SEMD). Functionally analogous to dual-energy material decomposition (DEMD)\cite{realvarezEnergyselectiveReconstructionsXray1976} for solid objects, this approach significantly extends the capabilities of grating-based dark-field CT. The SEMD framework enables the simultaneous reconstruction of absorption, phase-contrast, and dark-field images, alongside basis material images. By endowing the system with pseudo-dual-energy capabilities, this method broadens XGI's diagnostic utility beyond pulmonary applications, facilitating the quantitative imaging of other anatomical structures.

\section{Methods}

\subsection{Imaging Principles of XGI}
\begin{figure}[htbp]
    \centering\includegraphics[width=\columnwidth]{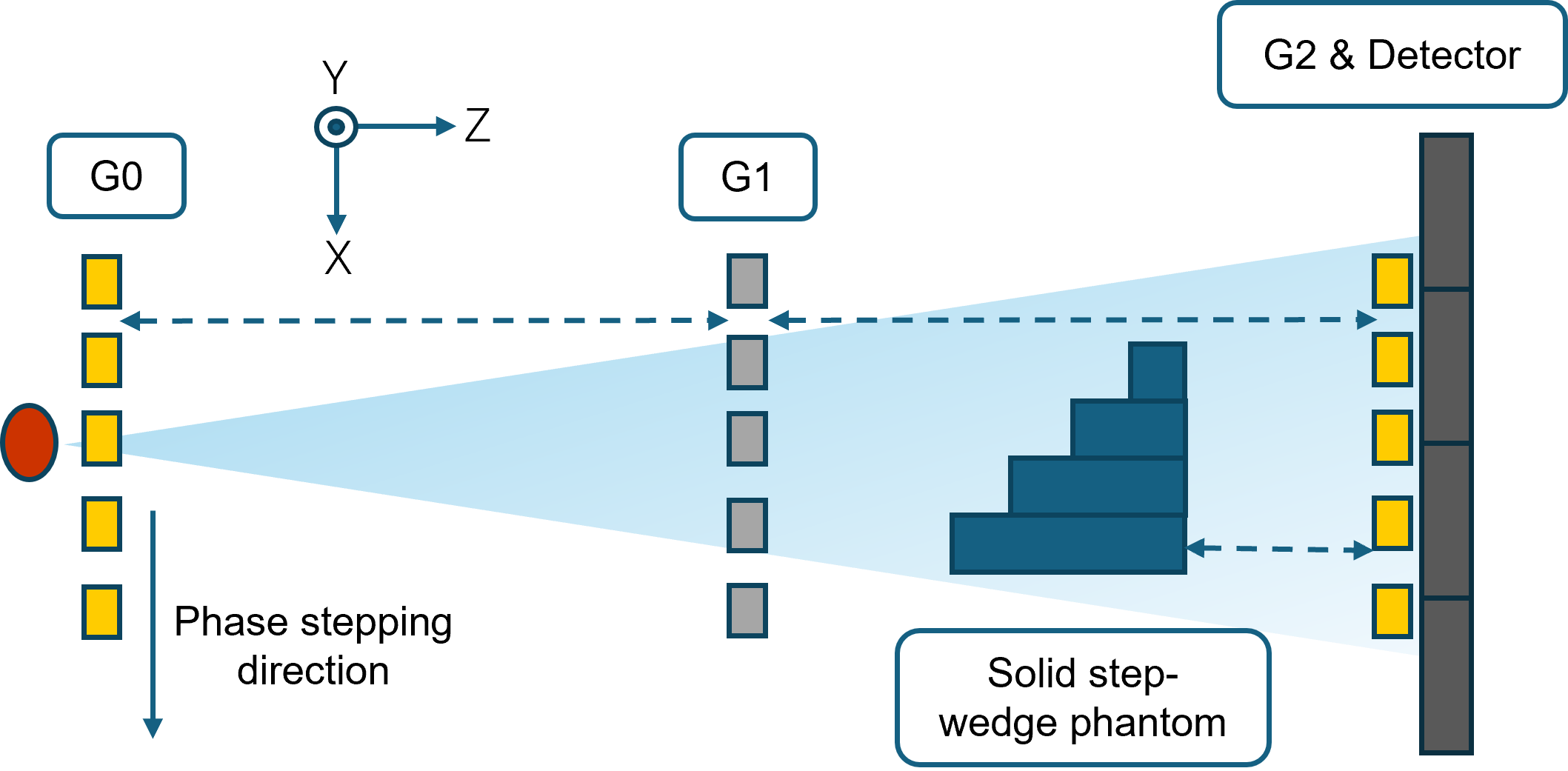}
    \caption{Schematic diagram of the symmetric Talbot-Lau interferometer. The step-wedge phantom is placed between the G1 and the G2 grating for spectrum estimation.}
    \label{fig:sys}
\end{figure}

In the grating-based dark-field imaging, X-ray intensity fringes are generated and analyzed using a Talbot-Lau interferometer consisting of three gratings as showing in Fig.~\ref{fig:sys}: a source grating G0, a phase grating G1, and an analyzer grating G2.
The G0 grating is positioned after the X-ray source to create an array of individually coherent but mutually incoherent line sources. The G1 grating generates periodic phase modulation on the wavefront, which leads to the formation of self-images at specific distances downstream, known as Talbot distances, which usually can not be directly detected by the detector. The G2 grating is placed at the position of the self-image to analyze the fringe pattern. The presence of a sample in the beam path causes changes in the fringe pattern, which can be captured by the detector and analyzed to extract absorption, phase-contrast, and dark-field information.

Phase stepping is the most commonly used and fundamental imaging scanning method. This method involves moving one of the gratings (usually G2) multiple times within a period along a direction perpendicular to the grating lines. After each step, the detector captures an image. Ultimately, a photon intensity curve that varies with the step position is obtained at each pixel, known as the phase-stepping curve. This curve is typically modeled as a first-order cosine function\cite{pfeifferHardXrayDarkfieldImaging2008}:
\begin{equation}
    I(\textbf{x};x_g,E) \simeq a_0(\textbf{x};E)+a_1(\textbf{x};E)\mathrm{cos}\left[\frac{2\pi}{p_2}x_g+\phi_1(\textbf{x};E)\right]
\end{equation}
where $\textbf{x}$ are the pixel coordinates on the detector plane, $x_g$ is the grating position during phase-stepping, $p_2$ is the period of G2. $a_i$ are the coefficients of the Fourier components and $\phi_1$ is the corresponding phase coefficient.

The visibility of the phase-stepping curve, $I(\textbf{x};x_g,E)$, is defined by the ratio:
\begin{equation}
    V(\textbf{x};E) = \frac{I_{max}-I_{min}}{I_{max}+I_{min}} = \frac{a_1(\textbf{x};E)}{a_0(\textbf{x};E)}
    \label{eq:visibility}
\end{equation}

If we denote the reference scan with the index $r$ and the sample scan with the index $s$, the transmission $T$ through the sample is calculated from the ratio of the mean values of the phase-stepping oscillation:
\begin{equation}
    T(\textbf{x};E) = \frac{a_{0,s}(\textbf{x};E)}{a_{0,r}(\textbf{x};E)} = \exp \left( -\int_L \mu(\textbf{r},E) dl \right)
    \label{eq:transmission}
\end{equation}
and the dark-field signal $D$ is defined as the visibility reduction:
\begin{equation}
    D(\textbf{x};E) = \frac{V_{s}(\textbf{x};E)}{V_{r}(\textbf{x};E)} = \exp \left( -\int_L \epsilon(\textbf{r},\xi) dl \right)
\end{equation}
where $\mu(\textbf{r},E)$ is the linear attenuation coefficient at position $\textbf{r}$ and the X-ray energy $E$. $\epsilon(\textbf{r},\xi)$ is the linear diffusion coefficient\cite{bechQuantitativeXrayDarkfield2010, wangQuantitativeGratingbasedXray2009} related to the auto-correlation length $\xi$\cite{stroblGeneralSolutionQuantitative2014a}, which is decided by system parameter and the X-ray energy $E$.

\subsection{Energy and Coupling Spectrum Estimation}
Under laboratory conditions, the spectral characteristics of the X-ray energy distribution $\Phi(E)$, together with the detector response $R(E)$, lead to a corresponding energy dependence of the fringe visibility $V(E)$. Consequently, both the measured intensity and visibility in the experiment are weighted sums over the relevant contributions\cite{demarcoXRayDarkFieldSignal2024}.

Unlike conventional imaging systems, during the phase-stepping process, the shape of the X-ray energy spectrum received by each pixel varies, $\Phi_{1}(E), \Phi_{2}(E), ..., \Phi_{N}(E)$.
At this point, the energy spectrum corresponding to the $a_0$ component of the phase-stepping curve is $\bar{\Phi}(E)$ (which is essentially the average of the energy spectrum at each step). For simplicity, this analysis focuses on a single pixel value, thereby omitting the spatial coordinates $\textbf{x}$, and the transmission $\bar{T}$ can be reformulated as follows:
\begin{equation}
    \bar{T} = \frac{a_{0,s}}{a_{0,r}} = \frac{\int_E \bar{\Phi}(E) \cdot T(E) dE}{\int_E \bar{\Phi}(E)dE}
    \label{eq:spec_T}
\end{equation}
where $a_0$ represents the mean value of the phase-stepping curve integrated over the entire energy range.

Similarly, the dark-field signal $\bar{D}$ can be expressed as\cite{demarcoXRayDarkFieldSignal2024}:
\begin{equation}
    \bar{D} = \frac{\bar{V}_s}{\bar{V}_r} = \frac{\int_E \overline{\Phi}(E) \cdot T(E) \cdot V(E) \cdot D(E) dE}{\int_E \overline{\Phi}(E) \cdot V(E)dE} \cdot \frac{1}{\bar{T}}
    \label{eq:spec_D}
\end{equation}

It can readily be observed that for a solid object without microstructures, where $D(E) = 1$, the beam hardening effect induced causes Eq.~\ref{eq:spec_D} to deviate from unity. The beam-hardening induced dark-field signal can be expressed as:
\begin{equation}
    \bar{D}_{H} =\frac{\int_E \overline{\Phi}(E) \cdot T(E) \cdot V(E) dE}{\int_E \overline{\Phi}(E) \cdot V(E)dE} \cdot \frac{1}{\bar{T}}
    \label{eq:spec_D_H}
\end{equation}

Given that the shape of the visibility spectrum $V(E)$ depends solely on the X-ray energy and system parameters (e.g., grating period, grating type etc.) and is independent of intensity, the energy spectrum and the visibility spectrum are mutually independent. They can therefore be combined into a single unified quantity, termed the coupling spectrum, defined as $\Omega(E) = \overline{\Phi}(E) \cdot V(E)$.
Substituting this definition into Eq.~\ref{eq:spec_D_H} and introducing an auxiliary variable $\bar{\Theta}$, the expression can be reformulated as:
\begin{equation}
    \bar{\Theta} = \bar{D}_{H} \cdot \bar{T} = \frac{\int_E \Omega(E) \cdot T(E) dE}{\int_E \Omega(E)dE}
    \label{eq:Theta2}
\end{equation}

Furthermore, considering the origin of the variable $\Theta$, which essentially represents the ratio of the first-order components of the phase-stepping curve. it can be derived from Eq.~\ref{eq:visibility} and \ref{eq:transmission}:
\begin{equation}
    \bar{\Theta} = \frac{\bar{V}_s}{\bar{V}_r} \cdot \frac{a_{0,s}}{a_{0,r}} = \frac{a_{1,s}}{a_{1,r}}
    \label{eq:Theta1}
\end{equation}

As established in previous studies on signal noise under polychromatic conditions, both the direct current component intensity $a_{0}$ and the first-order component intensity $a_{1}$ of the phase-stepping curve can be approximated as following Gaussian distributions under high signal-to-noise ratio (SNR) conditions\cite{chabiorSignaltonoiseRatioRay2011}:
\begin{equation}
    \begin{split}
        a_0 &\sim \mathcal{N}\left(\bar{a}_0, \frac{\bar{a}_0}{N}\right), \quad \text{SNR}_0 = \sqrt{N \bar{a}_0}
        \\
        a_1 &\sim \mathcal{N}\left(\bar{a}_1, \frac{2\bar{a}_0}{N}\right), \quad \text{SNR}_1 = \frac{\bar{V}}{2} \cdot \sqrt{N \bar{a}_0}
    \end{split}
    \label{eq:distributions}
\end{equation}
where $N$ is the number of phase steps, and $\bar{a}_0$ and $\bar{a}_1$ are the mean values of $a_0$ and $a_1$, respectively. $\mathrm{SNR}$ represents the signal-to-noise ratio of the corresponding component.

Eq.~\ref{eq:Theta2} shares an identical mathematical form with Eq.~\ref{eq:spec_T}, along with a consistent noise model, albeit with different noise levels. This structural equivalence enables spectrum estimation methods well-established in conventional absorption imaging to be directly extended to grating interferometry via the intermediate variable $\bar{\Theta}$.
By utilizing the transmission measurements acquired from solid samples of varying materials and thicknesses, both the energy spectrum $\bar{\Phi}(E)$ and the coupling spectrum $\Omega(E)$ can be independently and robustly estimated. For the theoretical validation of this study, solid step-wedge phantoms were employed, as schematically illustrated in Fig.~\ref{fig:sys}. Subsequently, the expectation-maximization (EM) algorithm was utilized to estimate both spectra, adhering to the iterative update scheme detailed in \cite{sidkyRobustMethodXray2005b}:
\begin{equation}
    w_{n}^{k+1}=\frac{w_{n}^k}{\sum_{m=1}^Ma_{mn}}\sum_{m=1}^M\frac{b_ma_{mn}}{\sum_{n'=1}^{N_s}w_{n'}^ka_{mn'}},\quad k=0,1,\ldots,
\end{equation}
with,
\begin{equation}
    a_{mn} = \exp [-\mu_m (E_n) L_m]
\end{equation}
where $M$ is the number of transmission measurements obtained from known calibration materials and $N_s$ energy bins are taken; $w_n$ is the normalized X-ray intensity at energy bin $E_n$;
$a_{mn}$ is the theoretical transmission of the $\mathrm{m}$-th material at energy bin $E_n$, and $b_m$ is the measured transmission that is integrated over the entire energy range, with $\mu_m (E)$ and $L_m$ being the material linear attenuation coefficients and thickness, respectively. The corresponding estimated transmission can be computed by $\sum_{n=1}^{N_s} w_n^k a_{mn}$.

Based on the estimated energy spectrum $\bar{\Phi}(E)$ and coupling spectrum $\Omega(E)$, the beam-hardening-induced dark-field signal for arbitrary materials can be theoretically computed by Eq.~\ref{eq:spec_D_H}. Then, this enables subsequent correction of the measured dark-field signal using established compensation methods\cite{tangPixelwiseBeamhardeningCorrection2023, viermetzInitialCharacterizationDarkField2023, lochschmidtLookupTablecorrectionBeam2025}.

\subsection{Single Energy Material Decomposition}
Building on the above analysis, it is further observed that the grating system inherently contains two independent spectra with acceptable energy separation, $\bar{\Phi}(E)$ and $\Omega(E)$.
This indicates that material decomposition can be achieved within a single imaging procedure by utilizing both $\bar{T}$ and $\bar{\Theta}$ (i.e. Eq.~\ref{eq:spec_T} and \ref{eq:Theta2}). For medical applications, the basis material decomposition method is typically used so that the linear attenuation coefficient of scanned object can be simplified as:
\begin{equation}
    \mu(\textbf{r},E) = \alpha_1(\textbf{r}) \cdot \mu_1(E) + \alpha_2(\textbf{r}) \cdot \mu_2(E)
\end{equation}
where $\mu_1(E)$ and $\mu_2(E)$ are the linear attenuation coefficients of two known basis materials.

The acquired intensity with sample can be expressed as:
\begin{equation}
    a_{i,s} = a_{i,r} \cdot \int_E S_i(E) \cdot e^{-A_1 \mu_1(E) - A_2 \mu_2(E)} dE
    \label{eq:SE_counts}
\end{equation}
\begin{equation}
    A_1 = \int_L \alpha_1(\textbf{r}) dl, \quad A_2 = \int_L \alpha_2(\textbf{r}) dl
\end{equation}
where $i = 0,1$ representing the two different normalized spectra. In grating interferometry, $S_0(E) = \bar{\Phi}(E)$ and $S_1(E) = \Omega(E)$, while in conventional dual-energy system, $S_0(E)$ and $S_1(E)$ represent the low and high energy spectra, respectively.

To evaluate the feasibility and theoretical performance of the proposed SEMD method, a comprehensive analysis is conducted from both qualitative and quantitative perspectives, with direct comparison against the DEMD method.

\begin{figure}[htbp]
    \centering\includegraphics[width=0.7\columnwidth]{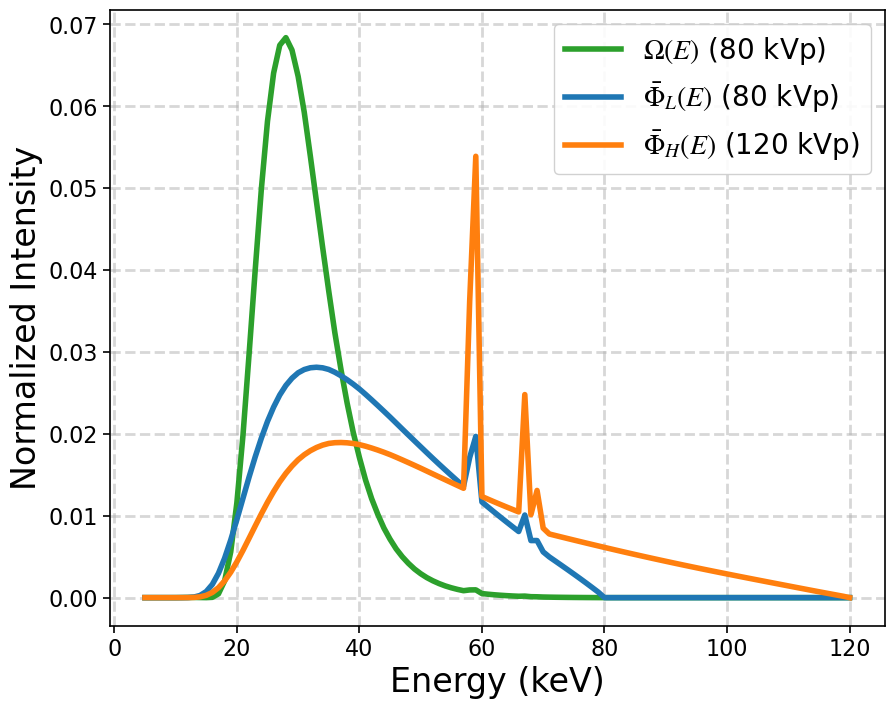}
    \caption{Area-normalized low and high energy X-ray spectra generated using SpekCalc\cite{poludniowskiSpekCalcProgramCalculate2009} with 2-$\mathrm{mm}$ Al filtration, assuming an ideal energy-integrating detector. The coupling spectrum $\Omega(E)$ of the grating interferometry, corresponding to the low energy configuration, was obtained via wave-optical simulation.}
    \label{fig:spec_shape}
\end{figure}

\subsubsection{Qualitative Performance Analysis}
For a given basis material and same background intensity.
Energy separation is a good indicator to roughly assess spectral imaging performance from a CT scan, which is defined as the difference between the effective energies of two spectra:
\begin{equation}
    \Delta E= \left| \frac{\int E\cdot S_0(E)dE}{\int S_0(E)dE}-\frac{\int E\cdot S_1(E)dE}{\int S_1(E)dE} \right|
\end{equation}
higher spectral separation is positively correlated with improved quality in basis material images.

We generated normalized low and high energy spectra at 80 kVp and 120 kVp for simulating dual-energy material decomposition, respectively, each filtered by 2 $\mathrm{mm}$-Al using SpekCalc software\cite{poludniowskiSpekCalcProgramCalculate2009}.
The visibility spectrum $V(E)$ was then obtained through a wave-optical simulation framework. Multiplying it by the $\bar{\Phi}_{L}(E)$ and normalizing the result yielded the coupling spectrum $\Omega(E)$. $\bar{\Phi}_{L}(E)$ and $\Omega(E)$ were subsequently utilized for the SEMD method. The three different spectra are shown in Fig.~\ref{fig:spec_shape}.
The spectral separation was calculated for each scenario, with the results showing that the dual-energy method achieved a spectral separation of 11.5 keV, while the SEMD method attained a separation of 10.9 keV.

In terms of spectral separation, the performance difference between the two methods is marginal. However, in the SEMD method, the intensity of $a_1$ is constrained by the visibility and is always lower than $a_0$ (e.g., under $\bar{\Phi}_{L}(E)$ condition, the visibility of phase-stepping curve was 0.148). Unlike the dual-energy method, the tube current cannot be adjusted to equalize the intensities of the two spectral components in the SEMD method. To further quantify the impact of this constraint, a detailed quantitative analysis is presented in the following.

\subsubsection{Quantitative Performance Analysis}
For material decomposition based on conventional dual-energy approach, the Cram\'{e}r-Rao lower bound (CRLB) can be employed to evaluate the theoretical performance limit\cite{roesslCramerRaoLower2009}. For the single energy material decomposition method, the CRLB analysis method remains the same, with the only difference lying in the form of the variance.
The intensity of the two acquired signals, $a_0$ and $a_1$, follow independent Gaussian distributions as mentioned earlier. Therefore, the joint probability distribution function of the acquired signals can be expressed as:
\begin{equation}
    P(a_{0,s}, ..., a_{i,s}) = \prod_{i} \frac{1}{\sqrt{2\pi \sigma_{i,s}^2}} \exp \left( -\frac{(a_{i,s} - \bar{a}_{i,s})^2}{2\sigma_{i,s}^2} \right)
\end{equation}

The negative log-likelihood function is then given by:
\begin{equation}
    \mathcal{L} = \sum_{i} \left( \frac{1}{2} \mathrm{ln}(2\pi \sigma_{i,s}^2) + \frac{(a_{i,s} - \bar{a}_{i,s})^2}{2\sigma_{i,s}^2} + C \right)
\end{equation}
Combining Eq.~\ref{eq:distributions} and \ref{eq:SE_counts}, the negative log-likelihood $\mathcal{L}$ now becomes a function of the $A$, and the Fisher information matrix can be calculated as:
\begin{equation}
    F_{mn} = \mathbb{E} \left[ \frac{\partial^2 \mathcal{L}}{\partial A_m \partial A_n} \right]
\end{equation}

The CRLB is given by the inverse of the Fisher information matrix, and the diagonal elements of the CRLB matrix provide the lower bounds on the variances of the estimated parameters:
\begin{equation}
    \mathrm{Var}(\hat{A}_m) \geq [\mathbf{F}^{-1}]_{mm}
\end{equation}

\begin{figure}[htbp]
    \centering
    \subfloat[\label{fig:CRLB1}]{\includegraphics[width=0.85\columnwidth]{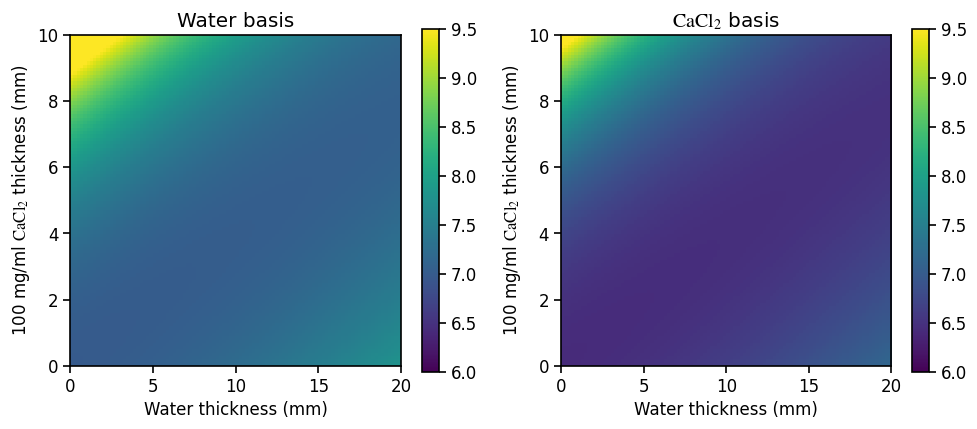}}
    \hfil
    \subfloat[\label{fig:CRLB2}]{\includegraphics[width=0.85\columnwidth]{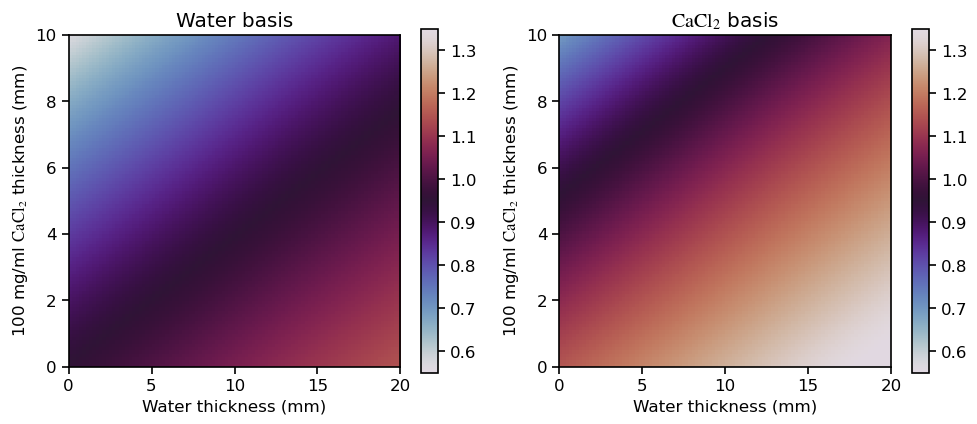}}

    \caption{(a) The ratio of the CRLB between the SEMD method at 80 kVp and the DEMD method at 80/120kVp, and (b) the ratio of the CRLB of the SEMD method at 120 kVp to that at 80 kVp, both evaluated for water and $\mathrm{CaCl_2}$ solution (100 mg/mL) across varying thickness combinations.}
    \label{fig:CRLB}
\end{figure}

To compare theoretical lower noise bounds between the dual-energy and the SEMD method.
The two basis materials selected were water and a $\mathrm{CaCl_2}$ solution (100 mg/mL). This choice matches the materials used in the subsequent wave-optical simulations and experimental validation.
To ensure a fair comparison, we kept the total photon number identical for both methods. For the SEMD method, the incident intensity was set to $1 \times 10^6$, while for the dual-energy method, the low and high energy intensities were set to $5 \times 10^5$ each. The results of CRLB ratio are presented in Fig.~\ref{fig:CRLB1}, which indicates that the noise level in the basis material images obtained with the SEMD method is markedly higher than that with the dual-energy method, by average value of approximately 7.3 and 6.7, respectively.

In the context of SEMD method, a higher spectral separation does not necessarily imply superior imaging performance. While increasing the tube voltage enhances spectral separation, it paradoxically degrades fringe visibility. This reduction attenuates the intensity of the first-order component, $a_1$, thereby amplifying the overall noise. As illustrated in Fig.~\ref{fig:CRLB2}, which depicts the ratio of CRLBs for SEMD-based basis material images at 120 kVp relative to 80 kVp, the average noise increases by factors of 1.15 and 1.37, respectively.
Despite the enhanced spectral separation at higher energies, the SEMD framework is inherently constrained by the diminishing signal strength of the $a_1$ component. Consequently, parameter selection should be governed by CRLB minimization based on the estimated spectra, rather than relying exclusively on spectral separation or fringe visibility as standalone metrics.

Crucially, the primary objective of the SEMD method is to maximize the data utilization of grating interferometry. The proposed framework not only yields the three standard imaging modalities but also provides basis material maps. The implementation protocol comprises three key steps: (1) Simultaneous estimation of the energy spectrum and the coupling spectrum using step-wedge phantoms; (2) Image-domain material decomposition following the reconstruction of $-\ln(\bar{T})$ and $-\ln(\bar{\Theta})$ projections acquired at an optimized tube voltage; and (3) Correction of beam-hardening-induced dark-field artifacts guided by the estimated spectra.

\section{Results}
\subsection{Experimental Setup}
An X-ray grating interferometry imaging platform was established for both wave-optical simulations and experiments. The grating parameters are detailed in Table~\ref{tab:Setup-parameters}. The system was configured for the first-order Talbot distance, with equal distance of approximately 14 cm between gratings G0 and G1, and between G1 and G2. In the wave-optical simulation, the source was placed directly adjacent to the G0 grating (i.e., zero distance). In the experiment, however, the distance between the source and G0 was set to 15 cm for larger field of view.
The experiments were conducted using a micro-focus X-ray tube (Hamamatsu L12161-07), which was operated in middle-focus mode. Images were recorded using an sCMOS camera (Gsense, PHOTONIC SCIENCE, UK) with a pixel size of $16.4 \times 16.4$ $\mathrm{\mu m^2}$ and a total resolution of $4096 \times 4096$ pixels and $8 \times 8$ binning was applied as the default setting. A phase-stepping protocol was adopted throughout the simulation and experimental work, with the specific parameters for individual measurements provided below. The Fourier analysis was applied to extract the $a_0$ and $a_1$ components from the acquisition phase-stepping curves\cite{pfeifferHardXrayDarkfieldImaging2008}.

\begin{table}[htbp]
    \caption{Key Grating Parameters of Our Imaging Platform}
    \label{tab:Setup-parameters}
    \centering
    \begin{tabular}{clcc}
        \hline
        \textbf{Grating} & \textbf{Type} & \textbf{Pitch ($\boldsymbol{\mathrm{\mu m}}$)} & \textbf{Duty Cycle} \\
        \hline
        $G_0$ & Absorption    & 4.8                     & 0.5                 \\
        $G_1$ & $\pi$-shift   & 4.8                     & 0.5                 \\
        $G_2$ & Absorption    & 4.8                     & 0.5                 \\
        \hline
    \end{tabular}
\end{table}

\begin{table}[htbp]
    \caption{Parameters of Step-Wedge Phantom}
    \label{tab:phantom-calibration}
    \centering
    \begin{tabular}{llcc}
        \hline
        \textbf{Usage} & \textbf{Material} & \textbf{Step Thickness (mm)} & \textbf{Total Steps} \\
        \hline
        \multirow{2}{*}{Calibration} & Al & 1 & 10 \\
                                     & PMMA & 3 & 10 \\
        \hline
        Validation                   & POM & 3 & 10 \\
        \hline
    \end{tabular}
\end{table}

\subsection{Energy and Coupling Spectrum Estimation}
To quantitatively assess the accuracy of the spectrum estimation in simulation, the normalized root mean square deviation (NRMSD), a standard evaluation metric in spectrum estimation, was used to measure the deviation between the estimated spectrum $w_i^e$ and the true spectrum $w_i^t$:
\begin{equation}
    \mathrm{NRMSD}=\frac{\sqrt{\sum_{i=1}^{N_s}(w_{i}^{e}-w_{i}^{t})^{2}/N_s}}{\max(w_{i}^{e})-\min(w_{i}^{e})}
\end{equation}

Additionally, the root mean square deviation (RMSD) between the estimated and measurement values of the transmission $\bar{T}$ and the variable $\bar{\Theta}$ was also calculated to assess the consistency between the estimated spectra and the measured projections:
\begin{equation}
    \mathrm{RMSD} = \sqrt{\frac{1}{N_m} \sum_{i=1}^{N_m} (x_i^e - x_i^m)^2}
\end{equation}

The normalized initial guess for the energy spectrum was defined as the simulated spectrum from a selected phase-stepping step after filtration by an Al filter of specified thickness. The initial guess for the coupling spectrum was then derived by multiplying this filtered spectrum with the simulated true visibility spectrum, followed by normalization. Identical initial guesses were employed in the subsequent simulations and experiments.

\subsubsection{Wave-Optical Simulations}
\label{sec:sim_spec}
A one-dimensional wave-optical simulation framework was used to validate the proposed spectrum estimation method.
The X-ray source was modeled as a polychromatic point source emitting an 80 kVp spectrum generated by the SpekCalc software\cite{poludniowskiSpekCalcProgramCalculate2009}, with the real detector response function of experimental detector incorporated.
For the phase-stepping curves, the background intensity of $a_0$ is set approximately to $3.3 \times 10^5$, and with a visibility of about 0.17, the corresponding first-order component intensity $a_1$ approximately $5.9 \times 10^4$.
The calibration phantoms used for spectrum estimation are presented in Table~\ref{tab:phantom-calibration}, which are identical to those employed in the experiments. All samples were located between the grating G1 and G2. A 5-step phase-stepping protocol was implemented.

\begin{figure}[htbp]
    \centering
    \subfloat[]{\includegraphics[width=0.48\columnwidth]{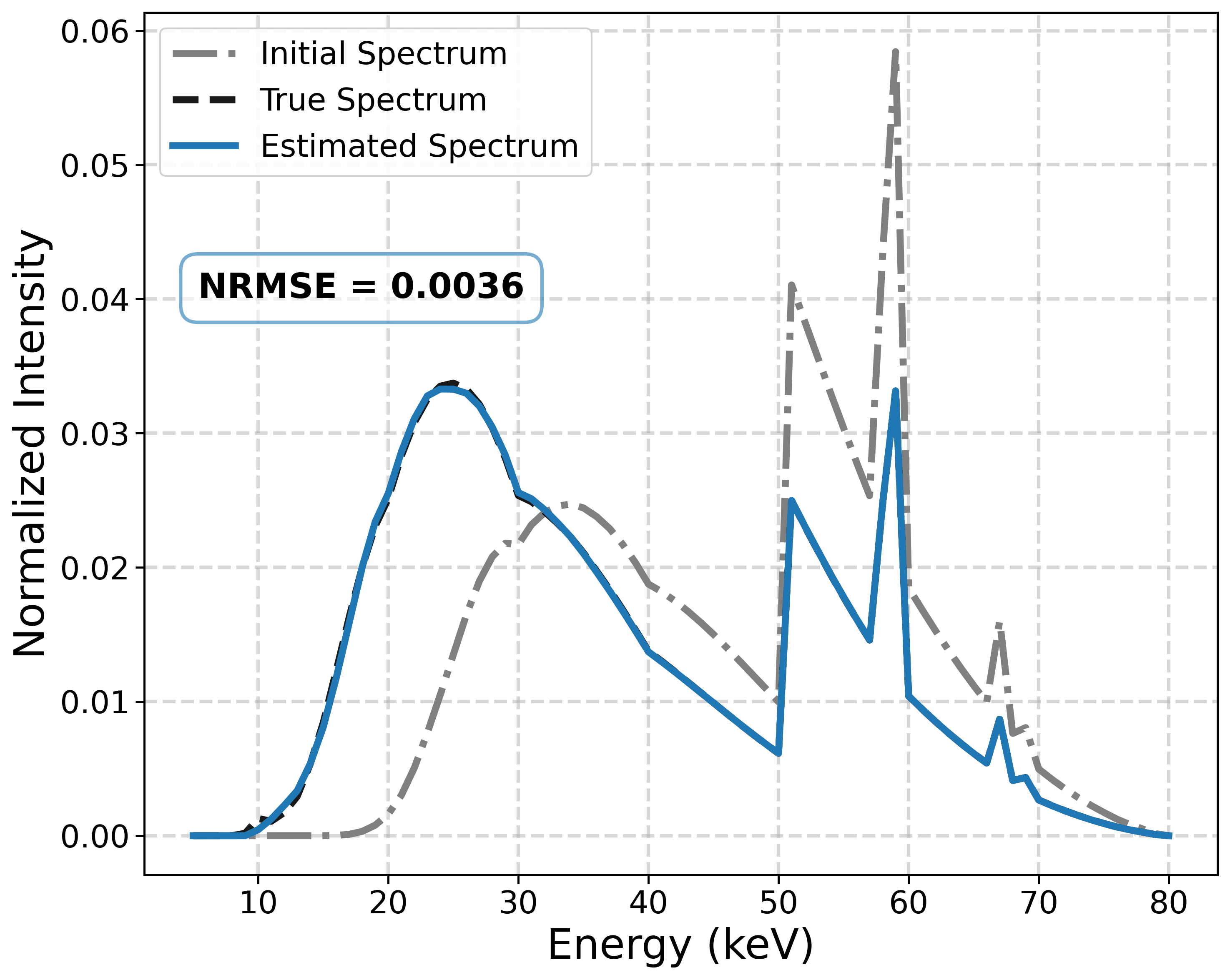}%
    \label{fig:sim_spec_withoutNoise1}}
    \hfil
    \subfloat[]{\includegraphics[width=0.48\columnwidth]{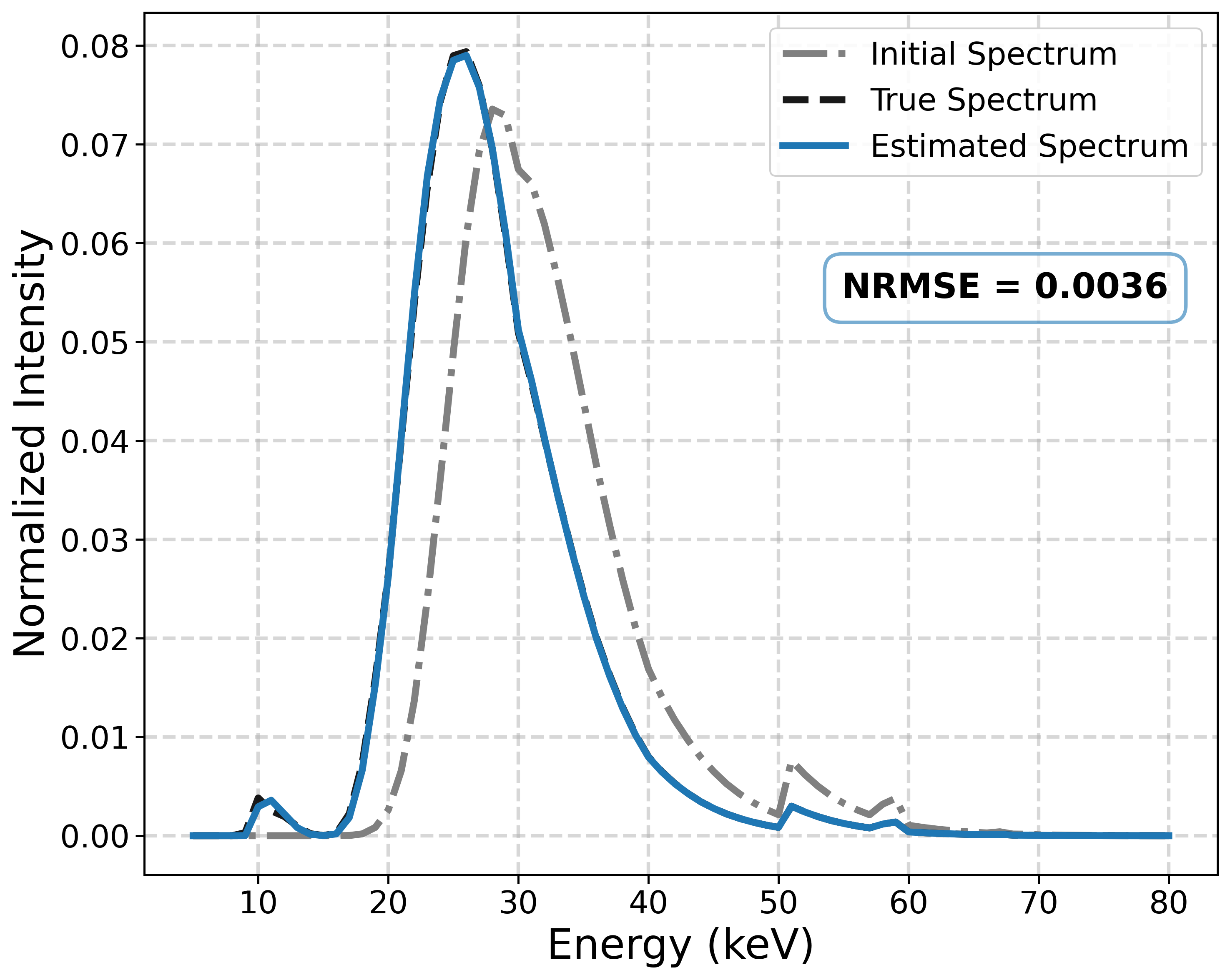}%
    \label{fig:sim_spec_withoutNoise2}}

    \subfloat[]{\includegraphics[width=0.48\columnwidth]{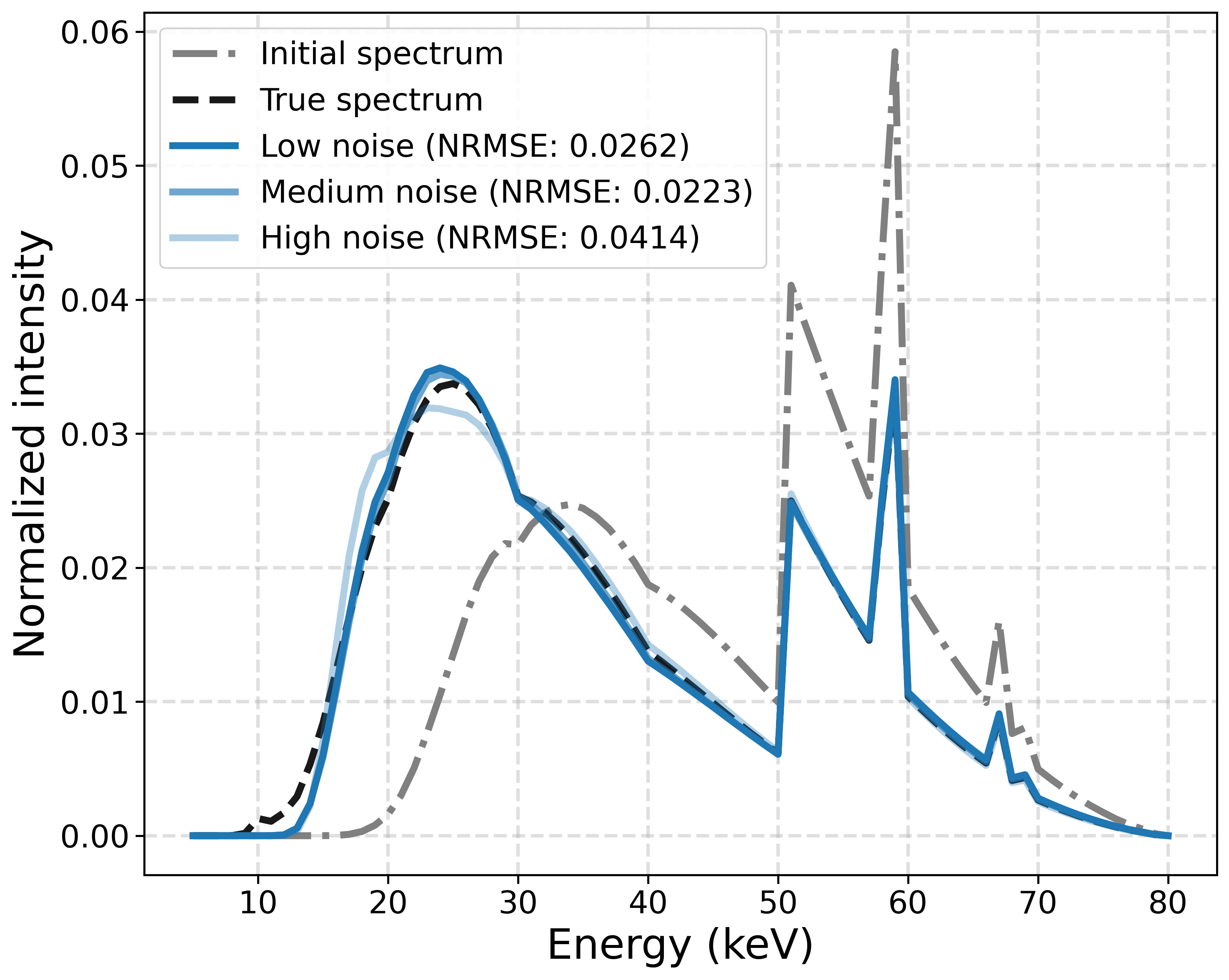}%
    \label{fig:sim_spec_withNoise1-1}}
    \hfil
    \subfloat[]{\includegraphics[width=0.48\columnwidth]{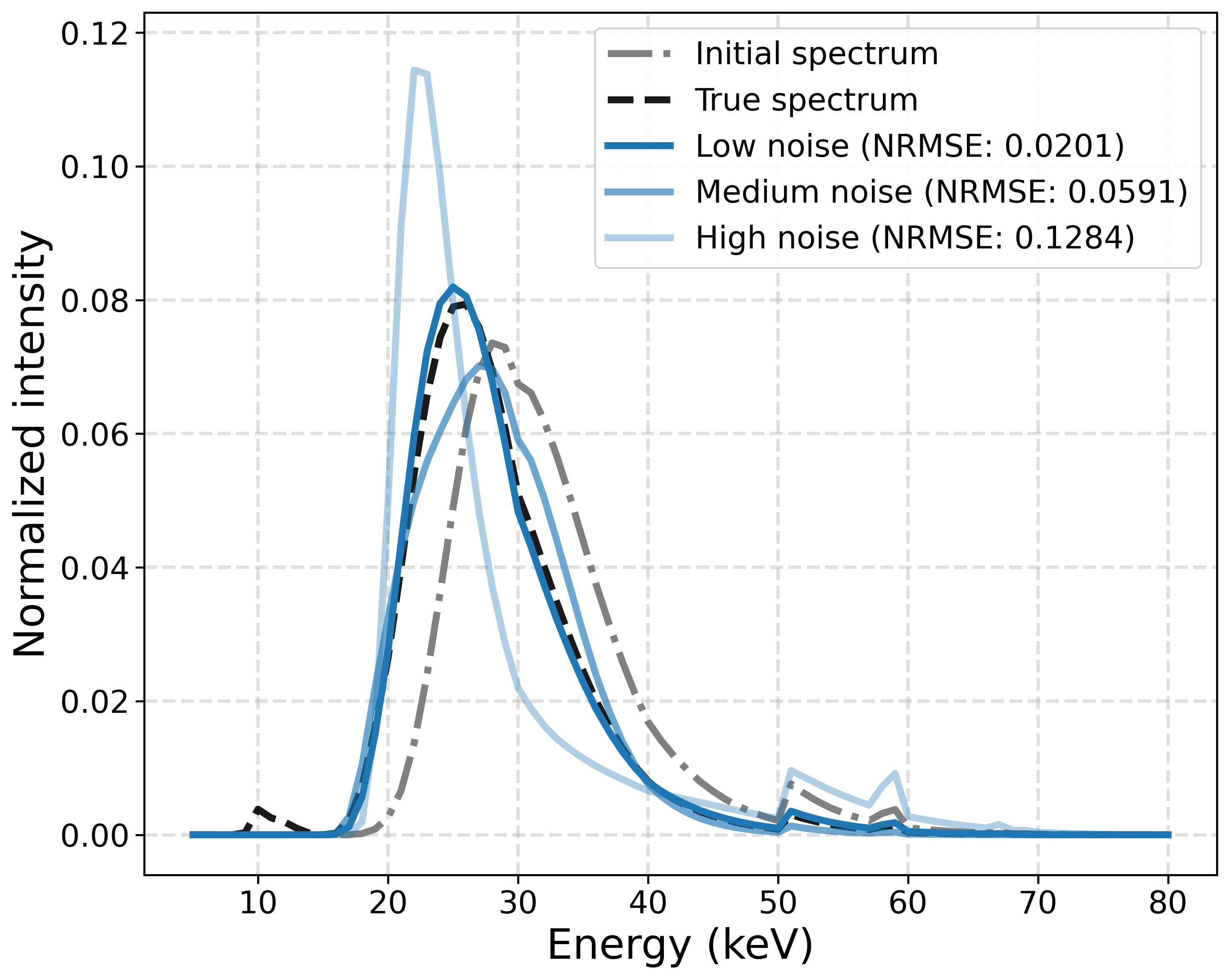}%
    \label{fig:sim_spec_withNoise1-2}}

    \caption{Simulation results of spectrum estimation under different conditions: (a) Estimated energy spectrum and (b) coupling spectrum in the absence of noise. (c) Estimated energy spectrum and (d) coupling spectrum under different noise levels.}
    \label{fig:sim_spec_noise}
\end{figure}

Initially, simulations were performed in the absence of noise to obtain signals from materials of varying types and thicknesses. The EM algorithm was then applied to separately estimate the energy spectrum and the coupling spectrum from the signals $\bar{T}$ and $\bar{\Theta}$, with the results shown in Fig.~\ref{fig:sim_spec_withoutNoise1} and \ref{fig:sim_spec_withoutNoise2}. The estimated spectra were then compared with the true spectra to evaluate the accuracy of the estimation method. The results show that under noise-free conditions, the estimated energy spectrum and coupling spectrum closely match their respective true spectra, thus demonstrating the validity of the method.

Subsequently, the influence of different noise levels on spectrum estimation was simulated and fixed same iteration number 2000 and 3000. Different levels of Gaussian noise were added to the intensity at each step of the phase-stepping curve and the corresponding results are presented in Fig.~\ref{fig:sim_spec_withNoise1-1} to \ref{fig:sim_spec_withNoise1-2}. The results indicate that the process of estimating the coupling spectrum is more sensitive to noise. Therefore, it is necessary to ensure that $\bar{\Theta}$ maintains an adequate SNR in experiments.

\begin{figure}[htbp]
    \centering
    \subfloat[\label{fig:sim_spec_withNoise2-1}]{\includegraphics[width=0.48\columnwidth]{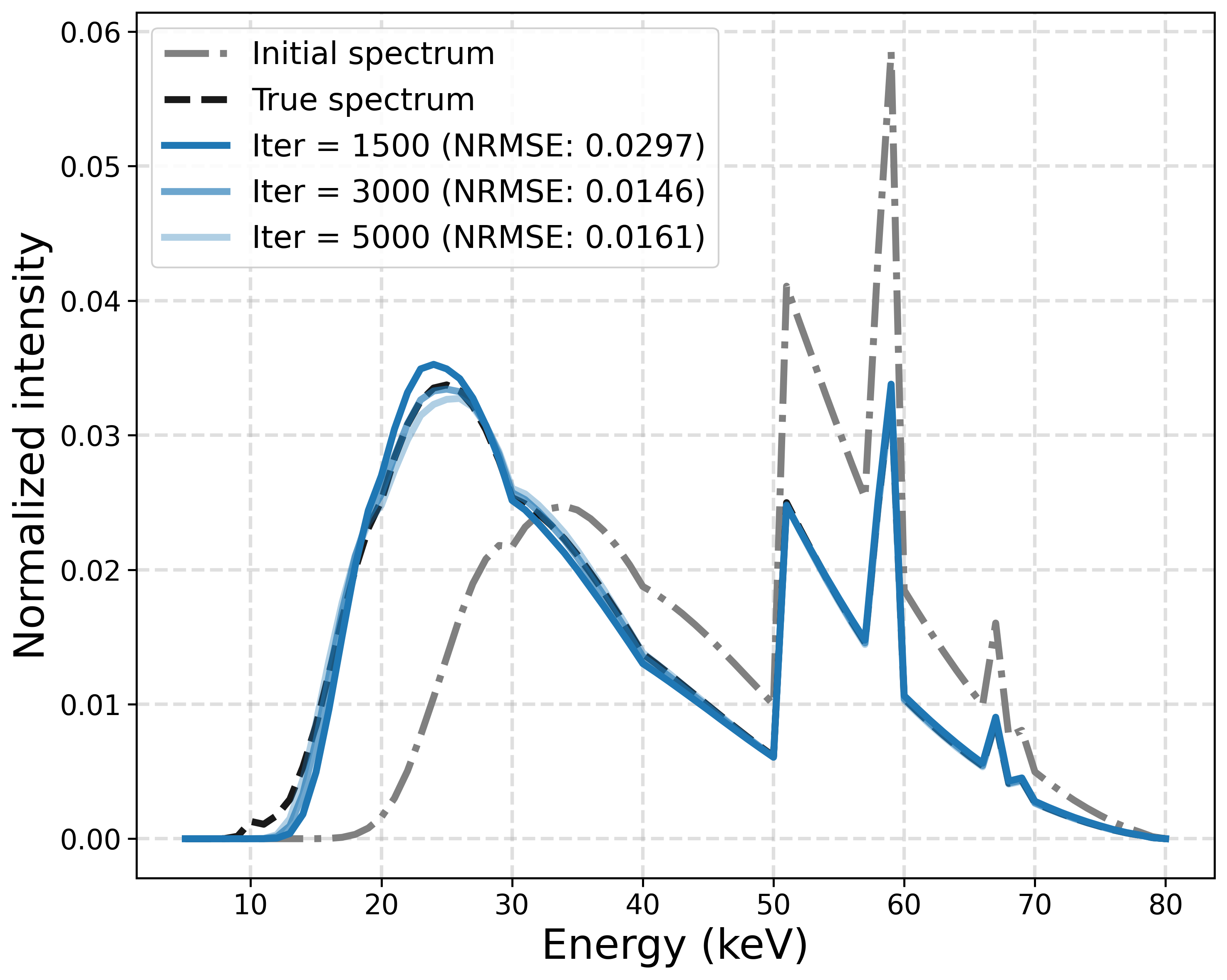}}%
    \hfil
    \subfloat[\label{fig:sim_spec_withNoise2-2}]{\includegraphics[width=0.48\columnwidth]{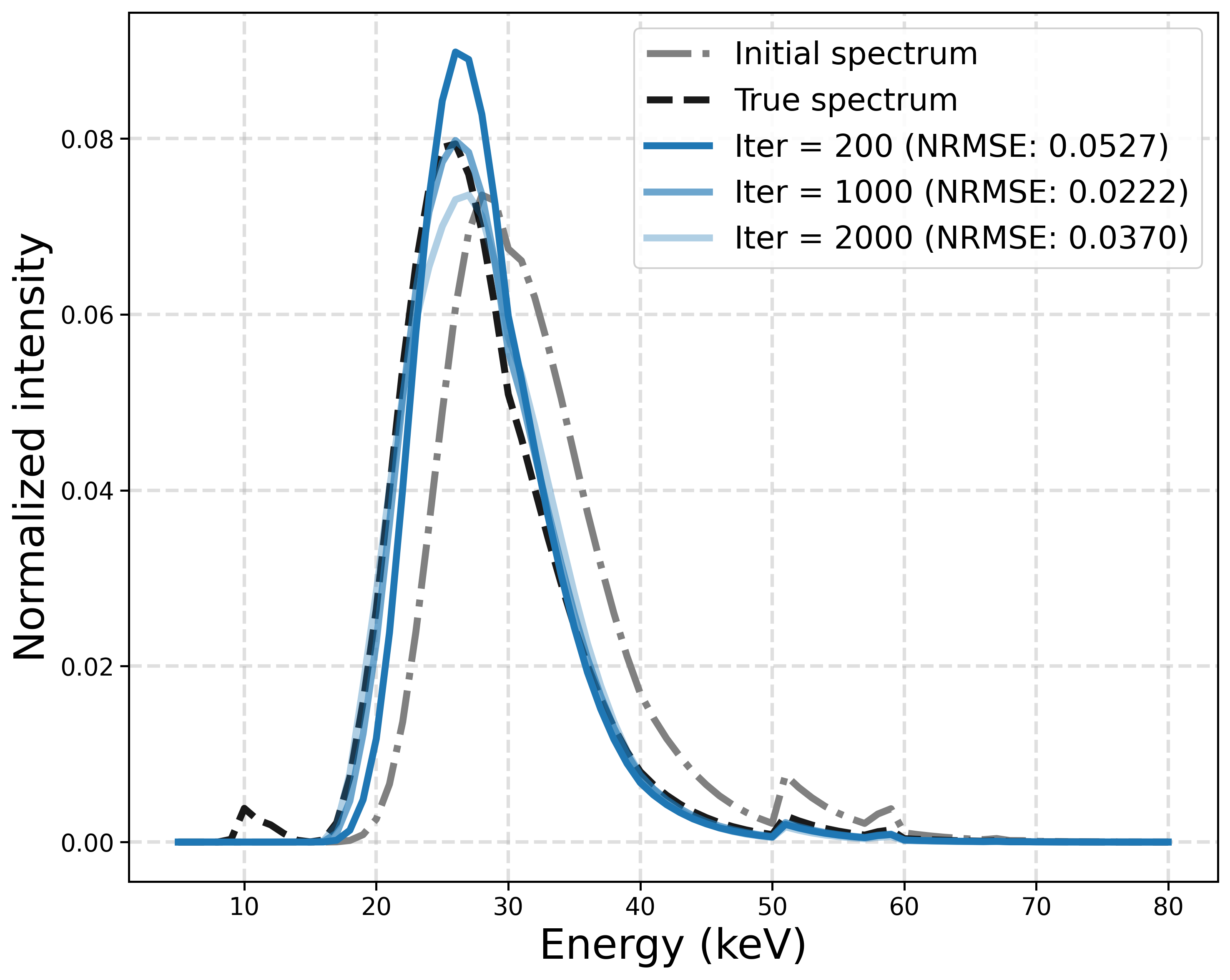}}%

    \medskip

    \subfloat[\label{fig:sim_spec_withNoise3-1}]{\includegraphics[width=0.48\columnwidth]{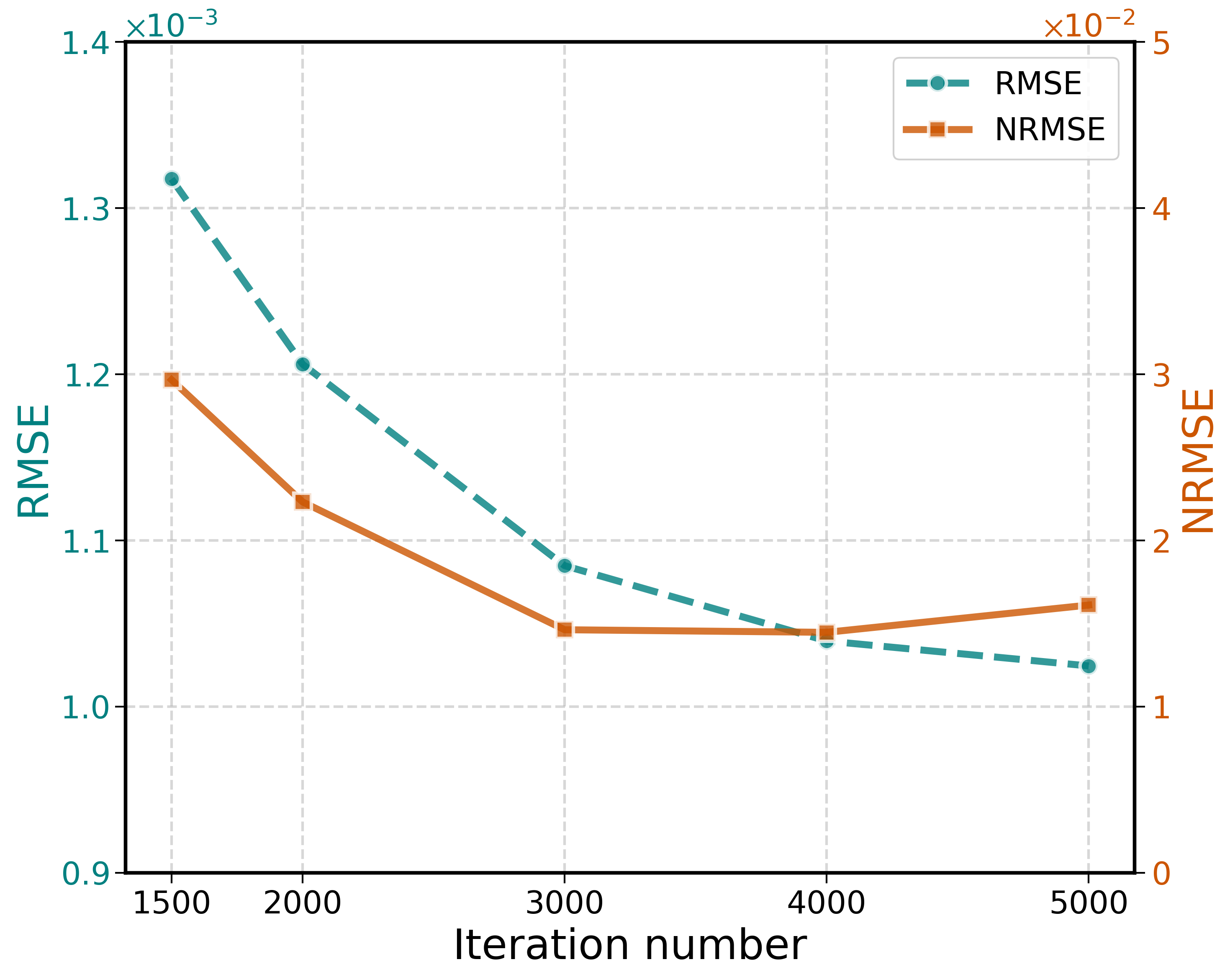}}%
    \hfil
    \subfloat[\label{fig:sim_spec_withNoise3-2}]{\includegraphics[width=0.48\columnwidth]{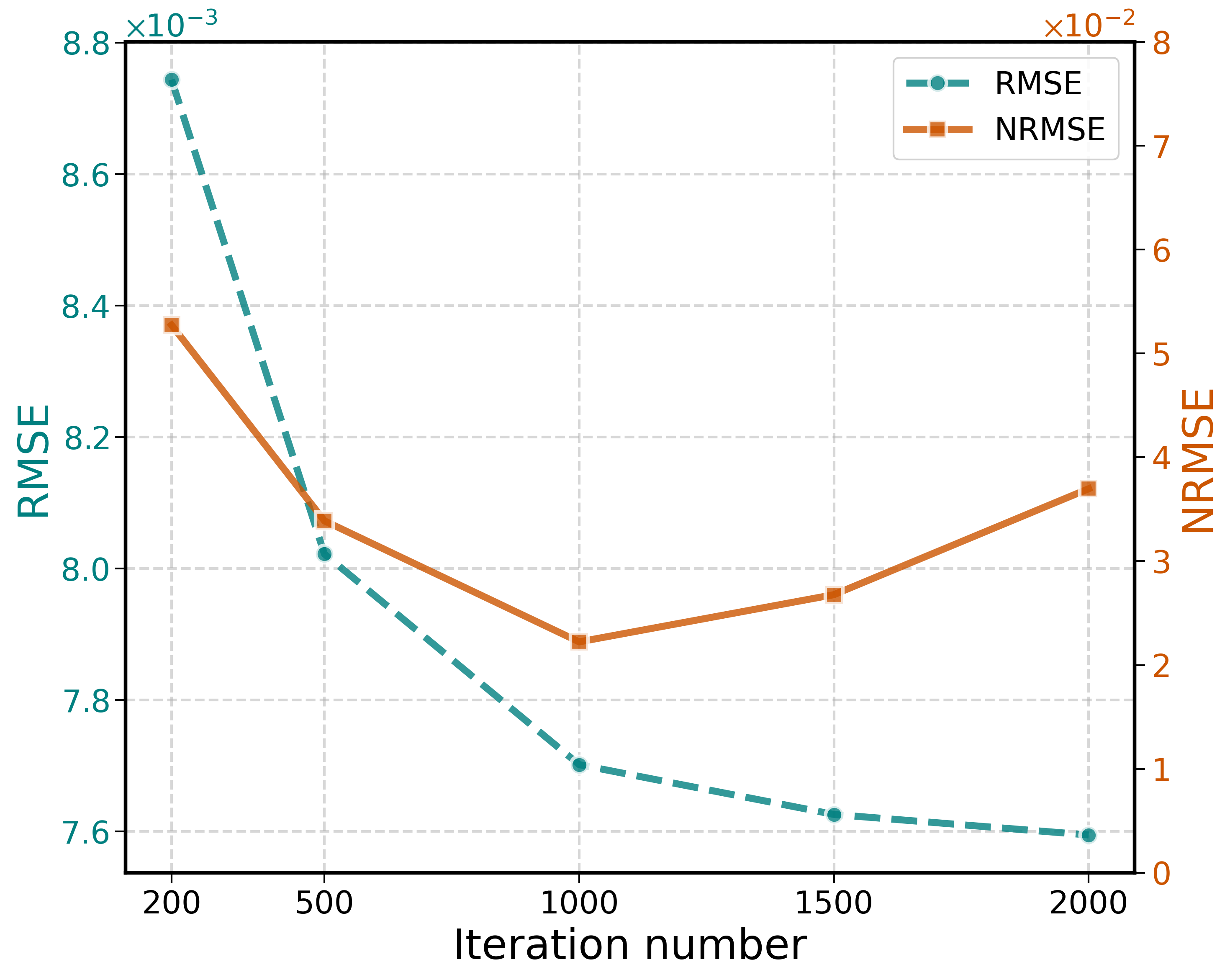}}%

    \caption{Simulation results of spectrum estimation under different iteration numbers: (a) Estimated energy spectrum and (b) coupling spectrum under different iteration numbers. (c) (d) The RMSE between the estimated and measured projections and the NRMSE between the estimated and true spectra versus iteration numbers.}
    \label{fig:sim_spec_iter}
\end{figure}

Finally, with the noise level maintained at a medium level, the number of iterations was varied. Fig.~\ref{fig:sim_spec_withNoise2-2} illustrates the impact of iteration count on the estimation of the coupling spectrum. Although a higher iteration count consistently reduces the RMSE---suggesting closer agreement between the estimated and measured projections---the NRMSE relative to the true spectrum does not decrease monotonically, as demonstrated in Fig.~\ref{fig:sim_spec_withNoise3-2}. This implies that early termination of the iteration is necessary. A similar effect was observed on the energy spectrum, Fig.~\ref{fig:sim_spec_withNoise3-1}.

\subsubsection{Experiments}
\label{sec:exp_spec}
\begin{figure}[htbp]
    \centering
    \subfloat[\label{fig:cail_phantom}]{%
        \includegraphics[width=0.48\columnwidth]{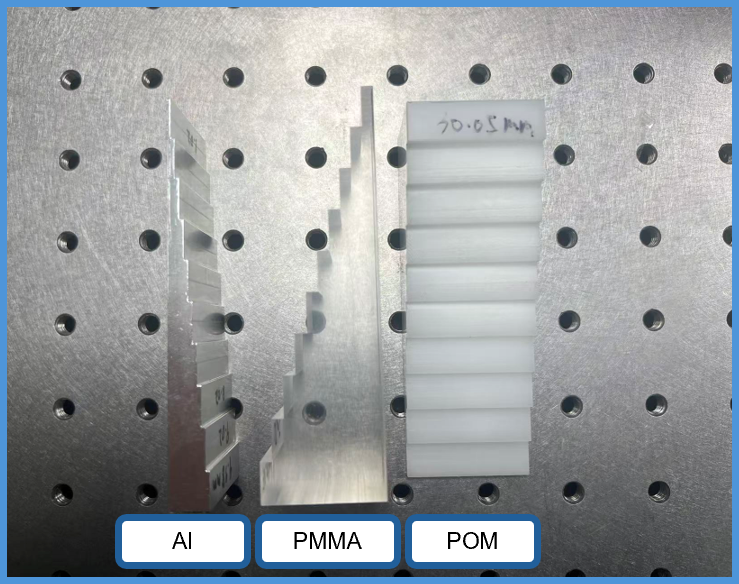}%
    }\hfil
    \subfloat[\label{fig:exp_spec_3}]{%
        \includegraphics[width=0.48\columnwidth]{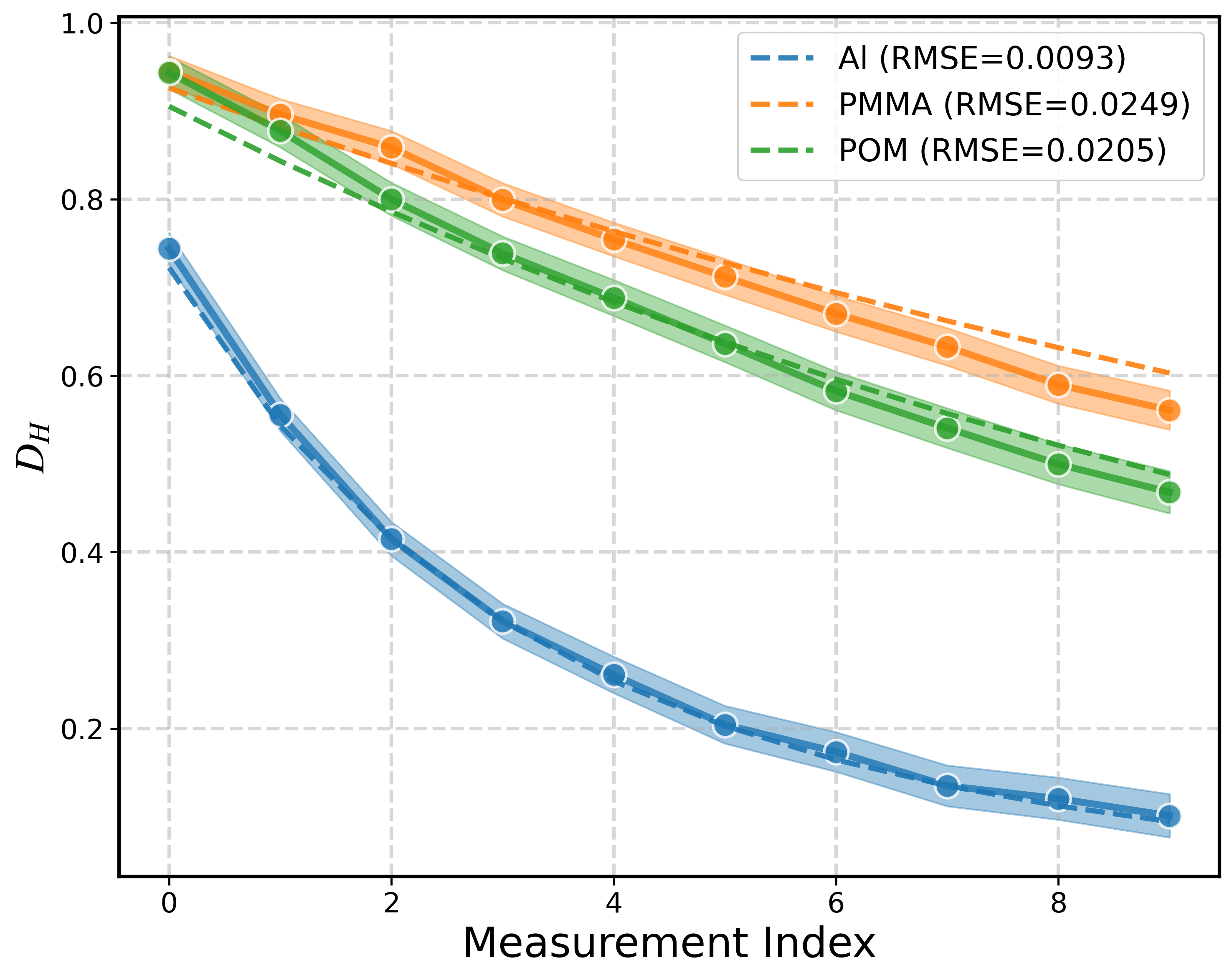}%
    }

    \medskip

    \subfloat[\label{fig:exp_spec_1-1}]{%
        \includegraphics[width=0.48\columnwidth]{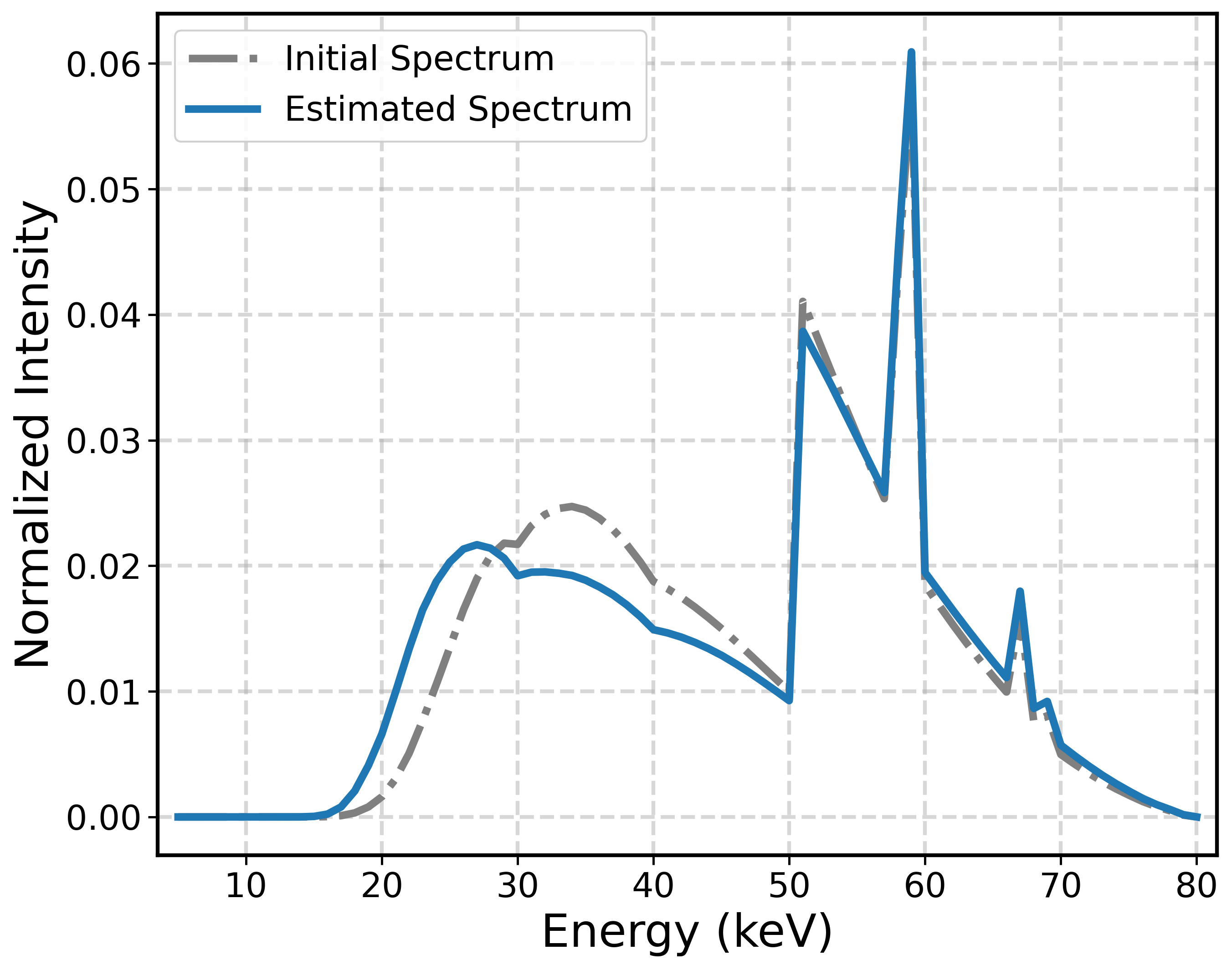}%
    }\hfil
    \subfloat[\label{fig:exp_spec_2-1}]{%
        \includegraphics[width=0.48\columnwidth]{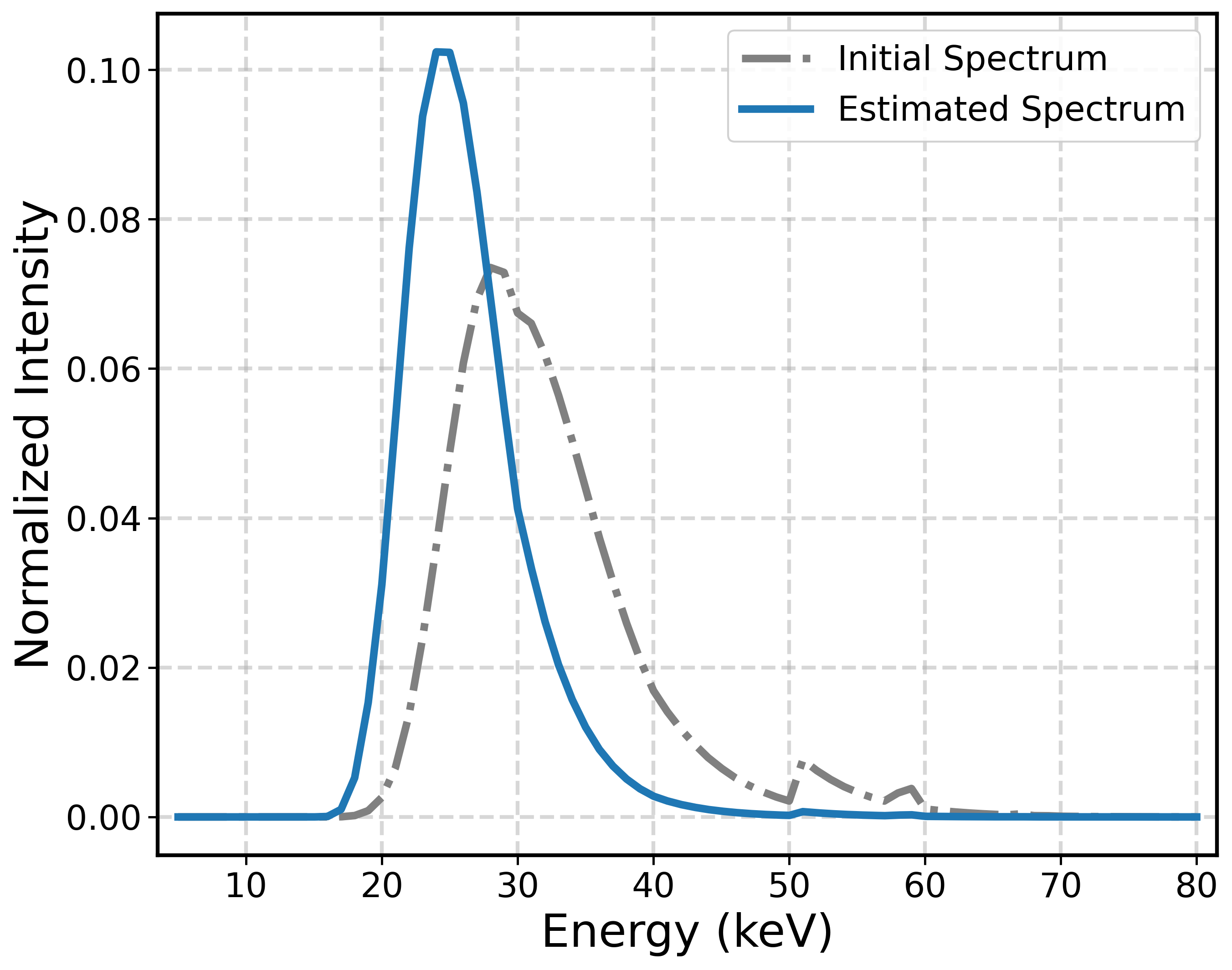}%
    }

    \caption{Experimental spectrum estimation results. (a) Photograph of the Al, PMMA, and POM step-wedge phantoms. (b) Measured and estimated beam-hardening-induced dark-field signals ($D_H$). (c) Estimated energy spectrum. (d) Estimated coupling spectrum.}
    \label{fig:experiment_results_spec}
\end{figure}

Since the ground truth is unavailable in experiments, an alternative test sample be employed to further evaluate the spectrum estimation results. POM (polyoxymethylene) was selected as the validation sample, and its measured data were excluded from the spectrum calibration process. The real step-wedge phantom used for calibration and validation is shown in Fig.~\ref{fig:cail_phantom}, with the specific parameters detailed in Table~\ref{tab:phantom-calibration}.
The step-wedge phantom was placed on a linear translation stage and sequentially imaged at the beam center for each thickness. Each imaging sequence consisted of a phase-stepping scan with the following parameters: 5 phase steps, 9 repeated acquisitions per step, and a 2-second exposure per frame to ensure enough SNR. The tube voltage was kept constant at 80 kVp and the tube current at 350 $\mathrm{\mu A}$.
For the measurement results, regions of interest (ROIs) were defined within the central optical path where intensity and visibility exhibited minimal variation, and the average values within these ROIs were calculated to represent the measured $\bar{T}$ and $\bar{\Theta}$ for each thickness.

\begin{figure}[htbp]
    \centering
    \subfloat[\label{fig:exp_spec_1-3}]{%
        \includegraphics[width=0.48\columnwidth]{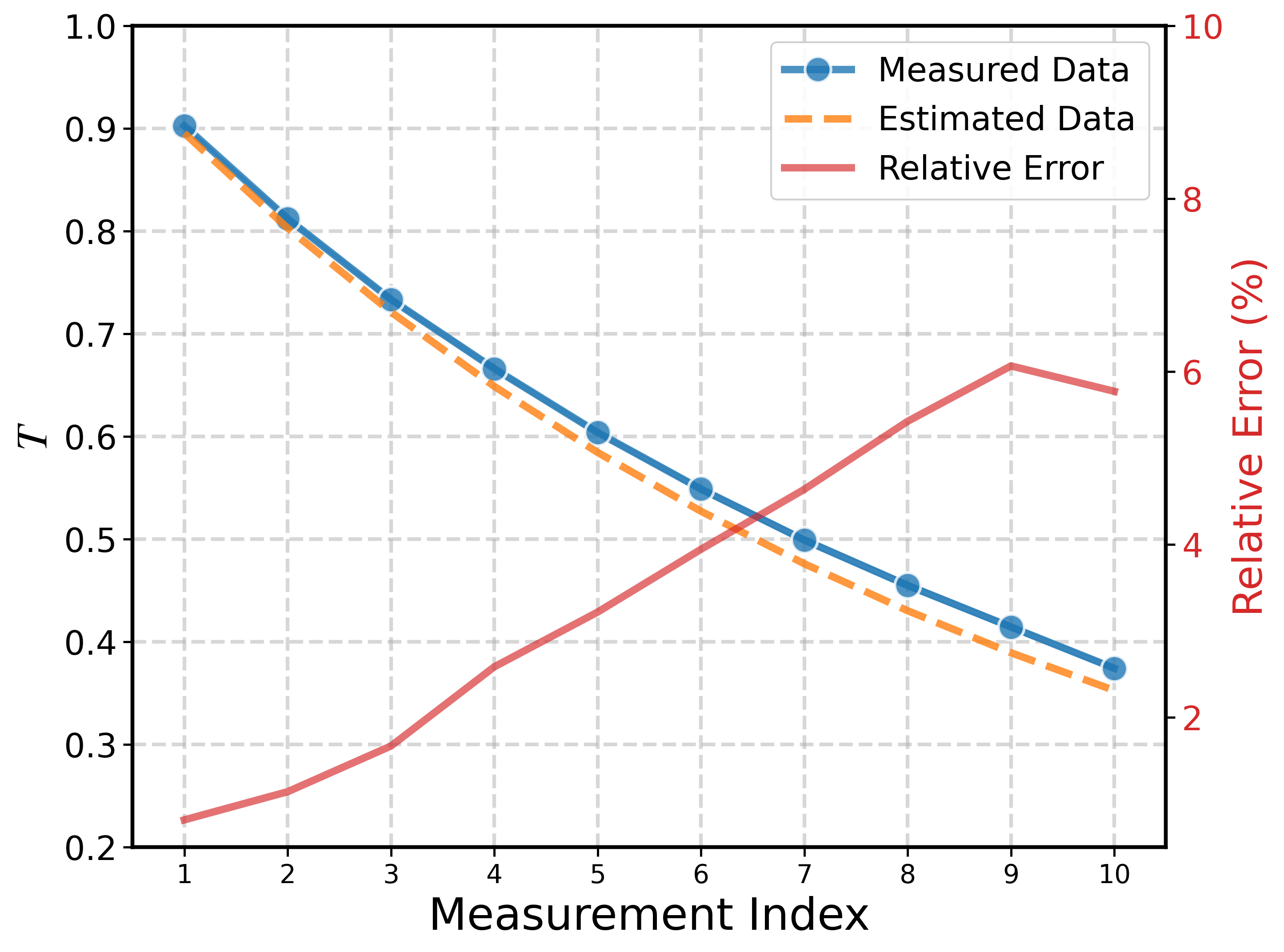}%
    }\hfill
    \subfloat[\label{fig:exp_spec_2-3}]{%
        \includegraphics[width=0.48\columnwidth]{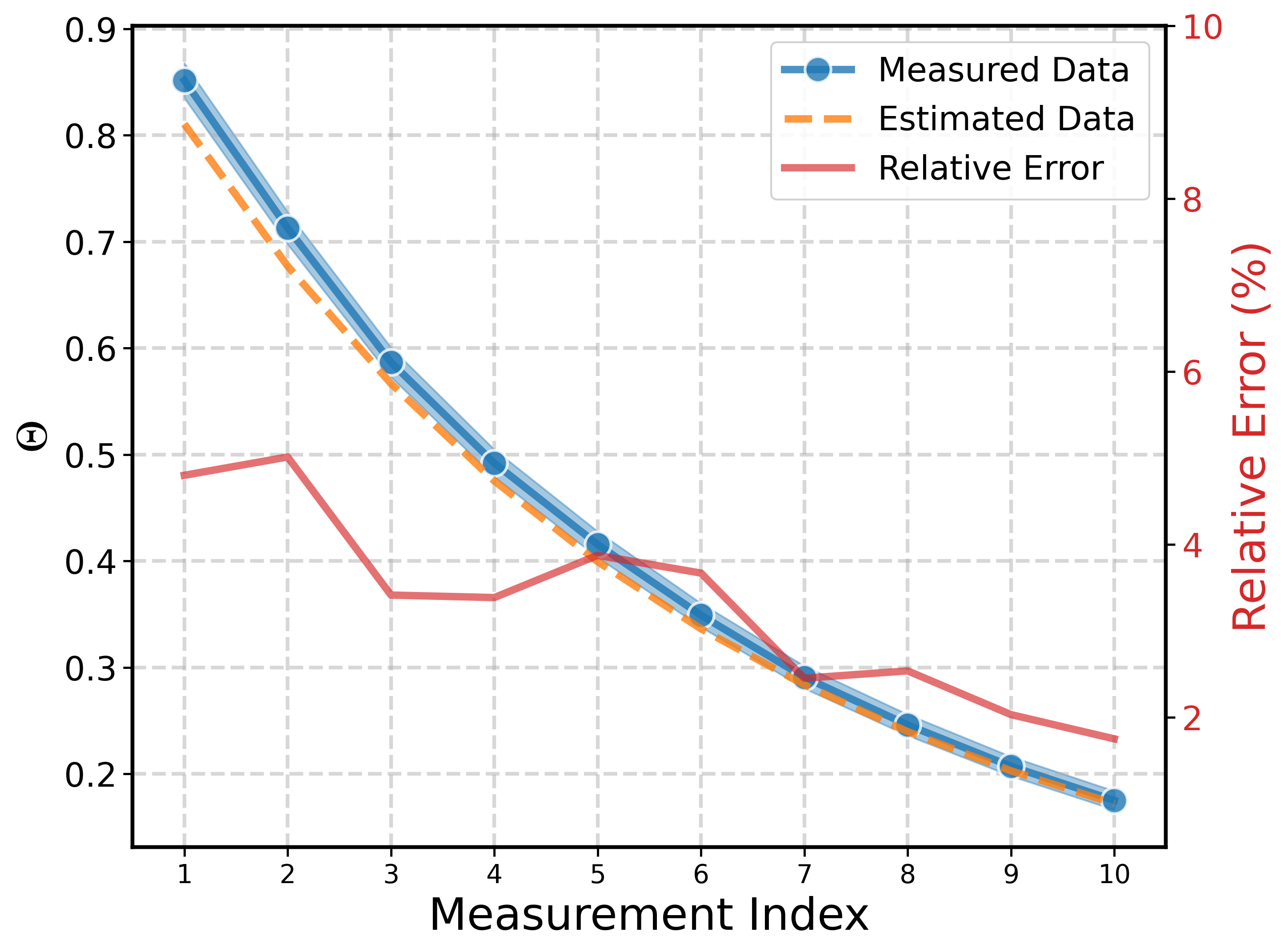}%
    }

    \caption{Validation of spectrum estimation on the POM step-wedge phantom. Measured and estimated values of (a) the $a_0$ transmission $\bar{T}$ and (b) the $a_1$ transmission $\bar{\Theta}$, with the corresponding relative errors (\%) overlaid and referenced to the right vertical axes.}
    \label{fig:estimation_err}
\end{figure}

The EM algorithm was subsequently applied to independently estimate the energy spectrum and the coupling spectrum from the measured signals $\bar{T}$ and $\bar{\Theta}$. Guided by prior simulation findings, the iteration counts were empirically optimized to 400 for the energy spectrum and 150 for the coupling spectrum. Both estimated spectra are depicted in Figs.~\ref{fig:exp_spec_1-1} and \ref{fig:exp_spec_2-1}, respectively. To quantitatively assess the fidelity of the estimation, we calculated the root-mean-square error (RMSE) between the measured and estimated projections. For the two calibration samples, this comparison yielded highly consistent results, achieving RMSEs of 0.0114 and 0.0088, respectively.
Furthermore, the beam-hardening-induced dark-field signal $\bar{D}_H$ for the entire phantom was computed utilizing the estimated spectra and compared against the measured values, as illustrated in Fig.~\ref{fig:exp_spec_3}.
Finally, Figs.~\ref{fig:exp_spec_1-3} and \ref{fig:exp_spec_2-3} present a comparison of the measured and estimated projection values for the validation sample, with each data point annotated by its relative error; notably, all errors remain strictly below 6\%.

Collectively, the results from both wave-optical simulations and experiments rigorously validate the proposed spectrum estimation method.
The high concordance between the estimated and true spectra in simulations, and the alignment of the measured projections and beam-hardening-induced dark-field signals in physical phantoms, substantiates the high fidelity and practical viability of the framework.

\subsection{Material Decomposition}
In this part, we quantitatively evaluate and compare the performance of the SEMD and DEMD frameworks via wave-optical simulations and experiments. The specific sample compositions are detailed in Table~\ref{tab:phantom-CT}.
To ensure a rigorous comparison, the DEMD configuration was established by solely removing the grating G2, while retaining G0 and G1 gratings. This strategy guarantees that the incident X-ray spectrum remains identical to that of the SEMD setup. Furthermore, experimental parameters were strictly controlled to maintain dose equivalence between the two methods. Consistent data processing was also applied, wherein the zeroth-order component $a_0$ for the DEMD dataset was extracted using the same Fourier analysis applied to virtual phase-stepping curves.

\begin{table}[htbp]
    \caption{List of the Materials and Concentrations. Concentrations are Indicated in the Parenthesis (Unit: mg/mL)}
    \label{tab:phantom-CT}
    \centering
    \begin{tabular}{ll|ll}
        \hline
        \textbf{Index} & \textbf{Material (concentration)} & \textbf{Index} & \textbf{Material (concentration)} \\
        \hline
        0 & EPDM                       & 3 & $\mathrm{CaCl_2}$ (50) \\
        1 & $\mathrm{CaCl_2}$ (100)    & 4 & $\mathrm{CaCl_2}$ (30) \\
        2 & $\mathrm{CaCl_2}$ (80)     & 5 & Water \\
        \hline
    \end{tabular}
\end{table}

\begin{figure}[htbp]
    \centering
    \subfloat[\label{fig:sim_decom1}]{\includegraphics[width=0.48\columnwidth]{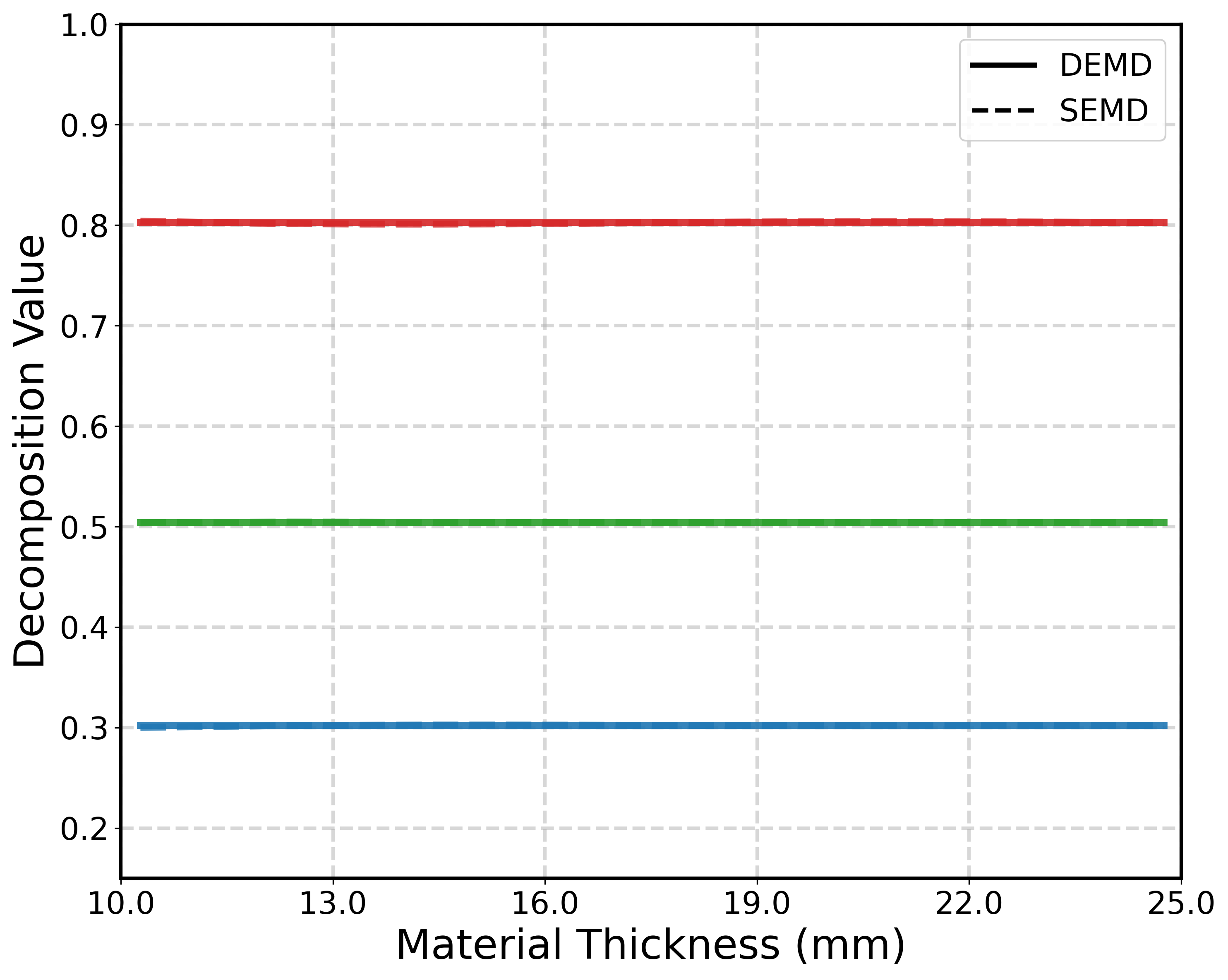}}%
    \hfil
    \subfloat[\label{fig:sim_decom2}]{\includegraphics[width=0.48\columnwidth]{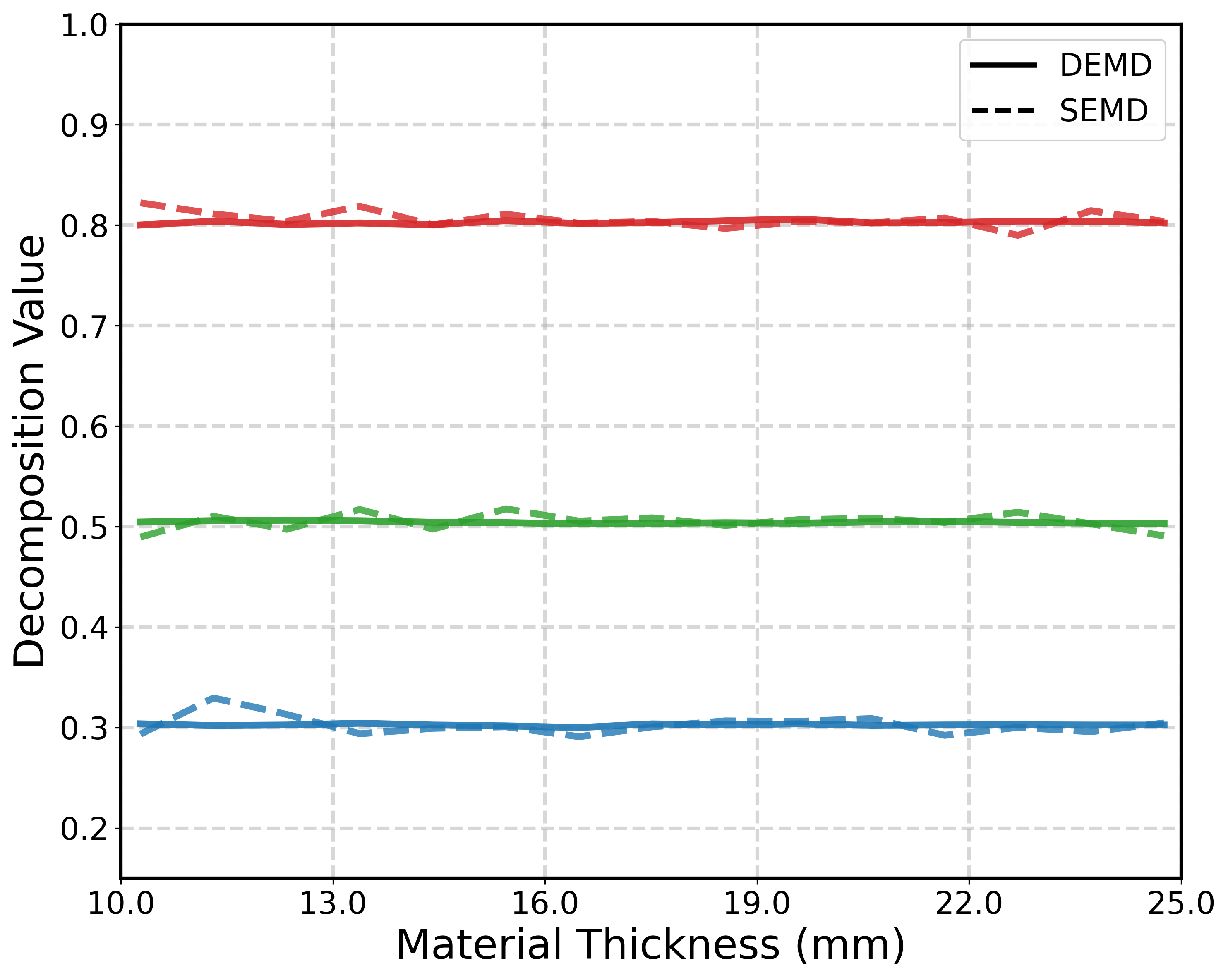}}%

    \medskip

    \begin{tabular}{lcc}
        \hline
        \textbf{GT} & \textbf{SEMD} & \textbf{DEMD} \\
        \hline
        0.3 & $0.3024 \pm 0.0096$ & $0.3026 \pm 0.0010$ \\
        0.5 & $0.5048 \pm 0.0082$ & $0.5044 \pm 0.0011$ \\
        0.8 & $0.8060 \pm 0.0081$ & $0.8028 \pm 0.0016$ \\
        \hline
    \end{tabular}

    \caption{Simulation results of two material decomposition method: (a) Noise-free results and (b) Noisy results. The table below shows the mean and standard deviation of the decomposed $\mathrm{CaCl_2}$ concentrations (unit: mg/mL) for both methods under noisy conditions.}
    \label{fig:sim_decom}
\end{figure}

\begin{figure}[htbp]
    \centering
    \includegraphics[width=\columnwidth]{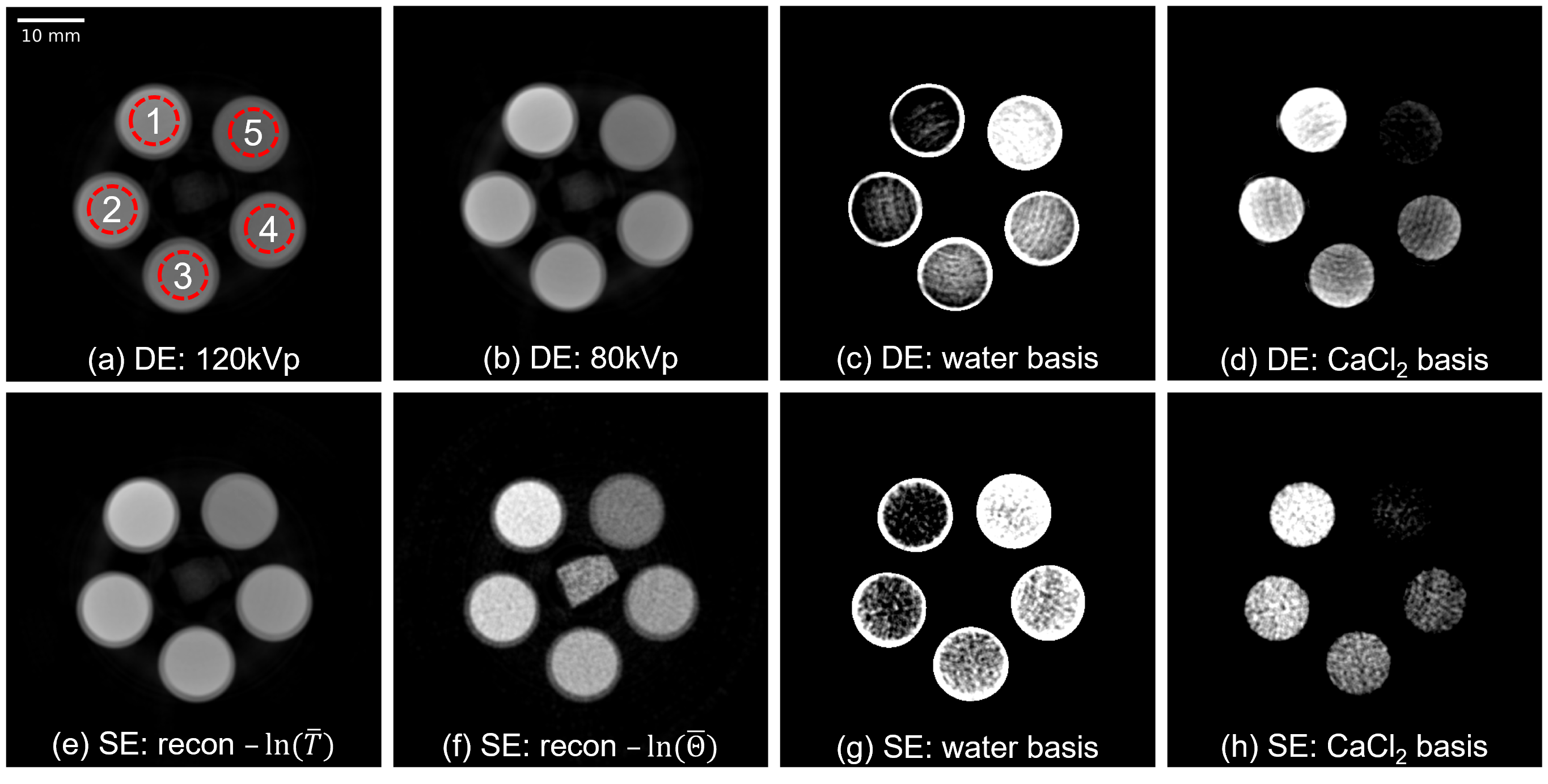}
    \caption{Experimental results of two material decomposition method. Image display windows: [0, 0.07] $\mathrm{mm^{-1}}$ for (a) and (b); [0, 0.05] $\mathrm{mm^{-1}}$ for (e); [0, 0.1] $\mathrm{mm^{-1}}$ for (f); and [0, 1] for all basis material image. The red dash circles indicate the ROIs used for quantitative analysis, and the scale bar in (a) represents 10 $\mathrm{mm}$.}
    \label{fig:exp_decom}
\end{figure}

\subsubsection{Wave-Optical Simulations}
\label{sec:sim_decom}
The SEMD simulations followed the parameter set described in Section~\ref{sec:sim_spec}. For DEMD, the same system layout was used except that grating G2 was omitted. The low and high energy acquisitions employed source spectra of 80 kVp and 120 kVp, respectively. The incident intensity was adjusted in the DEMD case to match the total photon number of the SEMD simulations.
Water and a $\mathrm{CaCl_2}$ solution (100 mg/mL) aqueous solution served as the basis materials. Step-wedge phantoms with a maximum thickness of 20 $\mathrm{mm}$ and 16 steps were created for each without container. A grid of signals was simulated for different thickness combinations, and a 6-order polynomial was inversely fitted to this data to establish the projection-domain material decomposition model\cite{chuangFastDualenergyComputational1987}.

For quantitative validation, aqueous $\mathrm{CaCl_2}$ solutions with concentrations of 30, 50, and 80 mg/mL were employed as test samples, corresponding to ground truth (GT) decomposition values of 0.3, 0.5, and 0.8, respectively. The signal values at various thicknesses were measured, substituted into the pre-fitted polynomial model for quantitative material decomposition.
Both noise-free and noisy conditions were simulated, with the results shown in Fig.~\ref{fig:sim_decom}. While the noise-free results verify the feasibility and performance parity of the SEMD method relative to DEMD, the noisy data reveal that SEMD yields higher noise levels at an identical dose. Nevertheless, its performance remains acceptable.

\subsubsection{Experiments}
To quantitatively evaluate the performance of the SEMD method, phantoms were constructed using aqueous $\mathrm{CaCl_2}$ solutions at varying concentrations together with water, which were sealed in plastic tubes with an outer diameter of 11 $\mathrm{mm}$ and an inner diameter of 10 $\mathrm{mm}$. The tubes were arranged with rotational symmetry on a dedicated holder, with an EPDM foam positioned at the center. The source-to-iso-center distance (SID) was set to 365 $\mathrm{mm}$, and the source-to-detector distance (SDD) was 430 $\mathrm{mm}$.
For SEMD experiment, a 5-step phase-stepping protocol was implemented, with 5 repeated acquisitions per step and a 2-second exposure per frame to ensure sufficient SNR. The tube voltage was maintained at 80 kVp and tube current was maintained at 350 $\mathrm{\mu A}$.
For DEMD experiment, gratings G0 and G1 were retained while only G2 was removed. The tube voltages were set to 80 kVp and 120 kVp for the low and high energy acquisitions, respectively. The tube current at 80 kVp was fixed at 350 $\mathrm{\mu A}$, while the current at 120 kVp was adjusted to 160 $\mathrm{\mu A}$ so that the background intensity were equalized. To ensure the identical sample dose, each acquisition involved $5\times5$ repeated exposures with a 1-second exposure per frame.

All projection data were acquired at 360 view angles over a full 360-degree rotation. Simultaneously, the object was translated out of the field of view at every 60-degree angular intervals to acquire background image sets.
For reconstruction, the simultaneous algebraic reconstruction technique (SART) was implemented using the TIGRE toolbox\cite{biguriTIGREV3Efficient2025}, running for 1000 iterations.
To mitigate artifacts arising from grating imperfections, a ring-artifact removal algorithm was applied in image-domain\cite{voSuperiorTechniquesEliminating2018}.
Following reconstruction, material decomposition was performed in the image domain via direct matrix inversion\cite{jiDualEnergyDifferential2020}.

\begin{figure}[htbp]
    \centering
    \subfloat[\label{fig:exp_water_compare}]{\includegraphics[width=0.48\columnwidth]{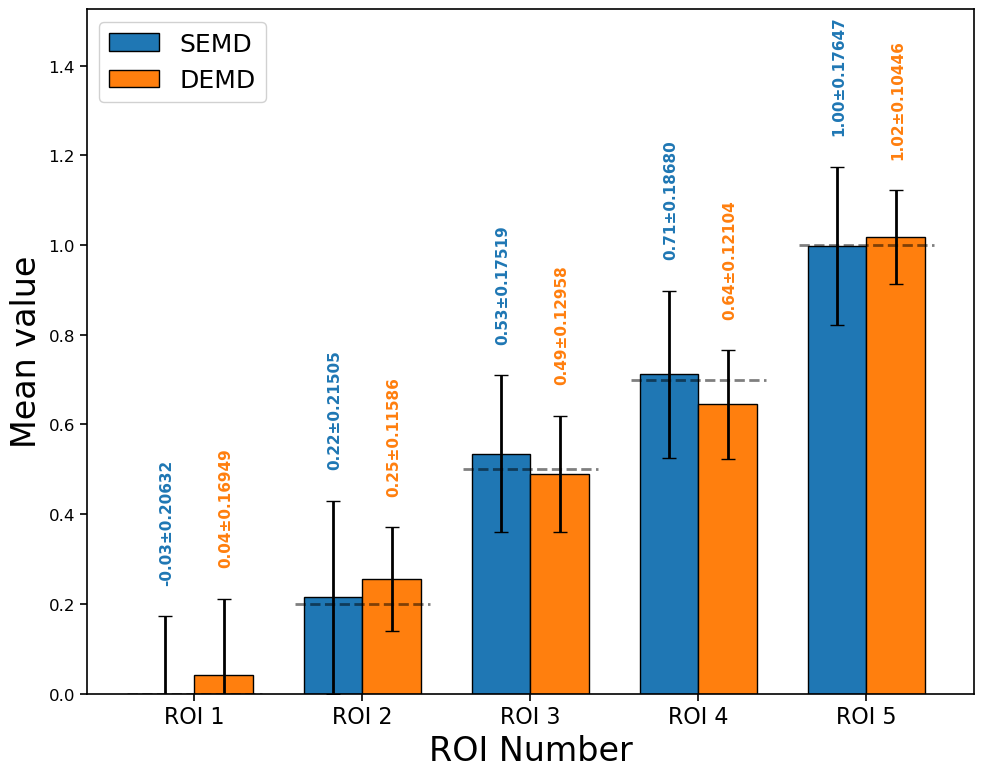}}%
    \hfil
    \subfloat[\label{fig:exp_CaCl2_compare}]{\includegraphics[width=0.48\columnwidth]{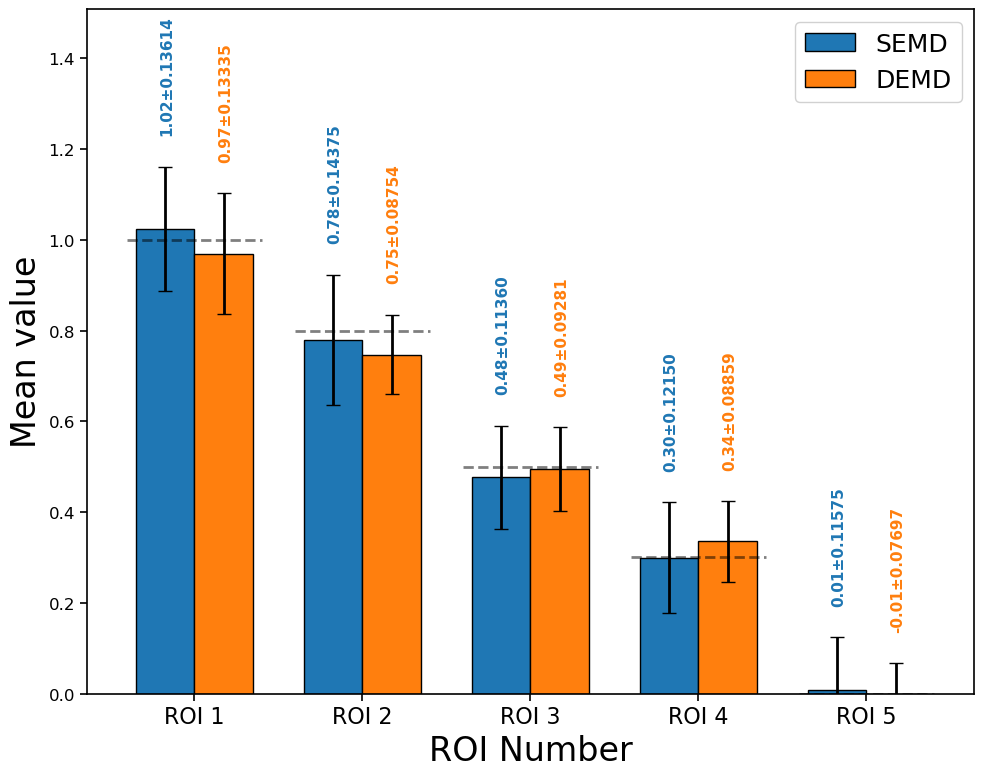}}%

    \caption{Comparison of the accuracy in basis material images obtained from both methods: (a) Water image and (b) $\mathrm{CaCl_2}$ (100 mg/mL) image. The dash lines represent the prepared concentrations. The dash lines represent the true concentrations.}
    \label{fig:exp_decom_compare}
\end{figure}

The basis material images with same slice using both methods are presented in Fig.~\ref{fig:exp_decom}. To improve the clarity of the basis material images, a mask derived from the absorption image was used to suppress artifacts in areas without the sample.
ROIs were defined within the tubes to calculate the mean and standard deviation of the decomposed water and $\mathrm{CaCl_2}$ concentrations for both methods, as summarized in the Fig.~\ref{fig:exp_decom_compare}. The results indicate that the mean values obtained from both methods closely align with the prepared concentrations, confirming the accuracy of the SEMD method. However, the standard deviations in the SEMD method are significantly higher than those in the DEMD method, which is consistent with the theoretical analysis and simulation results presented earlier. Meanwhile, the additional beam hardening caused by grating G2 in the SEMD setup results in a harder X-ray spectrum, leading to a noticeable reduction in cupping artifacts in the derived basis material images.

\begin{figure}[htbp]
    \centering
    \subfloat[\label{fig:exp_DF_corr}]{\includegraphics[width=0.42\columnwidth]{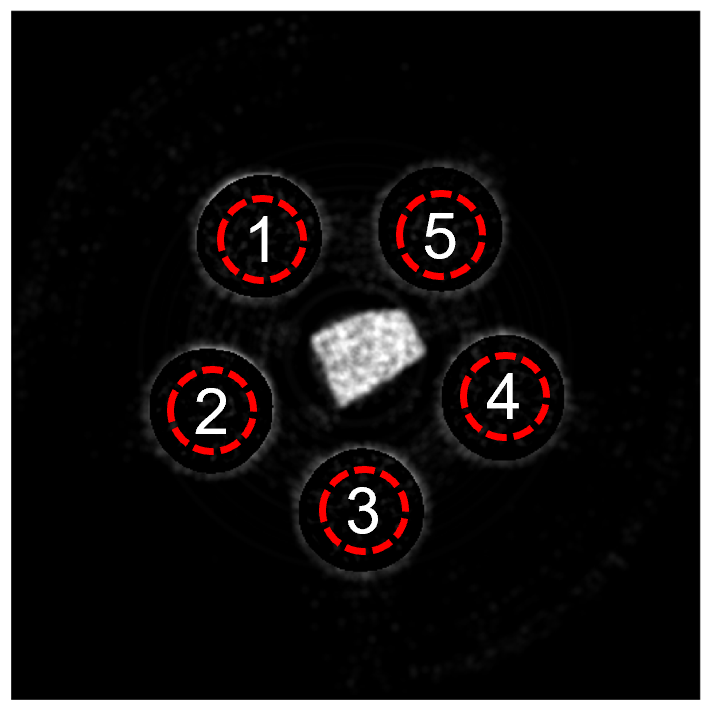}}%
    \hfil
    \subfloat[\label{fig:exp_DF_compare}]{\includegraphics[width=0.54\columnwidth]{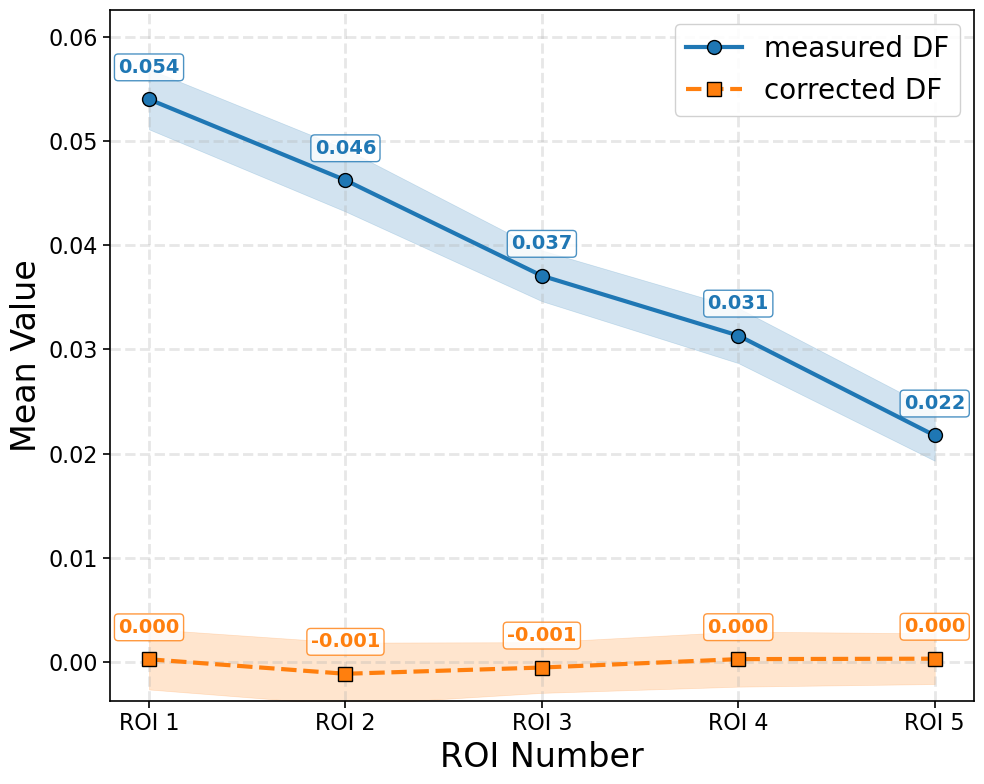}}%

    \caption{Evaluation of dark-field correction using estimated spectra: (a) Dark-field image after correction. (b) Comparison between the measured and estimated beam-hardening-induced dark-field signals for each material.}
    \label{fig:exp_DF_corr_compare}
\end{figure}

Furthermore, using the spectra estimated in Section~\ref{sec:exp_spec}, we evaluated the effectiveness of beam-hardening correction on dark-field images. The resultant corrected dark-field image of the entire phantom is presented in Fig.~\ref{fig:exp_DF_corr}.
A comparison between the measured and estimated beam-hardening-induced dark-field signals for each material is presented in Fig.~\ref{fig:exp_DF_compare}. The results demonstrate that the estimated values closely align with the measured data, confirming the validity of the spectrum estimation method for dark-field correction.
It is notable that while the weakly absorbing EPDM foam is nearly undetectable in the DEMD results, it presents pronounced contrast in the dark-field image.

\section{Discussion}
In this study, we identified an intrinsic coupling between the X-ray energy spectrum and the coupling spectrum in XGI. Leveraging this observation, we demonstrated the feasibility of simultaneously estimating both spectra using an EM algorithm and validated the approach through wave-optical simulations and experimental studies. A primary application of the retrieved spectra is the mitigation of beam-hardening-induced artifacts in dark-field images, which was demonstrated to be highly effective by the experimental results presented in Fig.~\ref{fig:exp_DF_corr_compare}.
Meanwhile, the estimated spectral information holds significant potential for correcting visibility-hardening in dark-field imaging \cite{demarcoXRayDarkFieldSignal2024}. Specifically, the estimated spectra can be employed to compute the thickness-dependent dark-field signal and derive the slope of its response curve. These data can then be used to fit a polynomial model that linearizes the hardened measurements, thereby enabling visibility-hardening correction. This process is conceptually analogous to the water beam-hardening correction widely adopted in conventional absorption imaging \cite{zhaoRobustBeamHardening2019}.

It should be noted that spectrum estimation is inherently an ill-posed inverse problem \cite{sidkyRobustMethodXray2005b}; consequently, the convergence of the solution is highly dependent on the choice of initial values. Therefore, the estimation fidelity could be further improved by employing more rigorous wave-optical simulations to generate the initial energy and coupling spectra. Additionally, the accuracy of the estimation framework can be enhanced by optimizing the design of the calibration phantom, specifically regarding its material composition and thickness combinations \cite{liEMEstimationXray2021}.
Finally, while the EM algorithm provides a robust mathematical foundation for this task, exploring alternative iterative reconstruction methods remains a valuable direction for future research \cite{zhaoIndirectTransmissionMeasurementbased2015, changSpectrumEstimationguidedIterative2020}.

Furthermore, we proposed the SEMD method, which leverages signals acquired from XGI to perform material decomposition for solid samples without microstrutures. This capability significantly expands the clinical applicability of XGI, particularly in the emerging field of dark-field CT.
Historically, the primary clinical focus of dark-field imaging has been restricted to the thorax, capitalizing on the strong ultra-small-angle scattering generated by pulmonary alveoli\cite{viermetzDarkfieldComputedTomography2022, guoOptimizationXrayDarkfield2025, spindlerSimulationStudyXray2025}. Conversely, its diagnostic utility in abdominal imaging has remained limited, as soft tissues lack such microscopic scattering interfaces and are instead dominated by pure absorption. By seamlessly integrating material decomposition into the XGI framework, our method simultaneously yields conventional absorption images, dark-field images, and quantitative basis material images. Consequently, even in abdominal regions devoid of strong dark-field signals, the system can still extract DEMD-equivalent diagnostic information (e.g., distinguishing iodine contrast agents from surrounding soft tissue). Substantiated by both wave-optical simulations and experiments, this integrated methodology demonstrates significant potential for transforming dark-field CT from a dedicated pulmonary scanner into a comprehensive, whole-body diagnostic modality.

One limitation of the that the proposed SEMD method lies in its underlying assumption regarding material properties. Specifically, projection-domain decomposition using SEMD is highly effective for solid objects lacking microstructures (i.e., pure-absorption materials), as corroborated by the simulation results in Section~\ref{sec:sim_decom}.
However, in typical clinical scenarios such as pulmonary CT, where strong scattering interfaces are present, this prerequisite is violated; consequently, the framework must rely exclusively on image-domain decomposition.

Moreover, under identical dose conditions, the SEMD method inherently yields basis material images with higher noise levels compared to the DEMD approach. This noise amplification stems from the intrinsically lower intensity of the first-order harmonic $a_1$, which is fundamentally bounded by the XGI's fringe visibility. The resulting reduction in the SNR of $a_1$ propagates directly into the decomposed basis images, particularly during image-domain decomposition. To mitigate this noise penalty, future implementations could integrate iterative image-domain decomposition techniques \cite{zhangSpectralCTImagedomain2023} alongside advanced deep learning based denoising method \cite{wangEmulatingLowdosePCCT2025}.
In the future, the integration of photon-counting detectors (PCDs) with grating interferometry systems represents a highly promising technological trajectory\cite{jiDualEnergyDifferential2020}. Importantly, the SEMD framework proposed in this study possesses intrinsic scalability; it can be readily extended to multi-energy material decomposition by leveraging data across multiple energy bins. Ultimately, this synergy between advanced detector hardware and our spectrum estimation methodology holds immense potential for further broadening the clinical and biomedical applications of XGI.

\section{Conclusion}
In conclusion, we identified an intrinsic similarity between the X-ray energy spectrum and the system-related coupling spectrum in XGI, and established an EM-based framework that enables their simultaneous estimation. The accuracy of this method was comprehensively validated through wave-optical simulations and experiments, and the retrieved spectra were demonstrated to effectively mitigate beam-hardening artifacts in dark-field imaging.
Building upon this spectral characterization, we introduced the SEMD method, which leverages XGI signals with single energy to achieve quantitative material decomposition. Substantiated by rigorous experimental and simulation data, this integrated methodology significantly broadens the application scope of grating interferometry, highlighting its immense potential for advancing diagnostic modalities, particularly dark-field CT.


\subsection*{Funding}
This study is supported in part by National Natural Science Foundation of China (No. 62227804), Beijing Natural Science Foundation (No. L232115) and National key research and development program of China (No. 2023YFC2605802).

\subsection*{Acknowledgment}
The authors thank all collaborators for their support in this work.

\subsection*{Disclosures}
The authors declare no conflicts of interest.

\subsection*{Data Availability Statement}
The data supporting the experimental results are available from the authors upon reasonable request.


\bibliographystyle{unsrt}
\bibliography{ref}

\end{document}